\documentclass[12pt,a4paper]{article}
\usepackage{xcolor,paralist,url}
\usepackage[round]{natbib}
\usepackage[pdftex,colorlinks=true,hypertexnames=false]{hyperref}
\definecolor{darkblue}{rgb}{0,0,.6}
\hypersetup{citecolor=darkblue,linkcolor=darkblue,urlcolor=darkblue}
\usepackage{amssymb,enumitem,booktabs,subfig,setspace,bm,longtable,xr,dsfont,orcidlink}
\usepackage[top=1in, bottom=1in, left=1in, right=1in]{geometry}
\usepackage{url}
\usepackage{amsmath}
\usepackage{amsfonts}
\usepackage{epsfig}
\usepackage{graphics}
\usepackage[normalem]{ulem}
\usepackage{palatino,eulervm}
\usepackage{graphicx}
\usepackage{lscape}
\usepackage{rotating}
\usepackage{multirow}
\usepackage{bm}  
\usepackage{booktabs}
\usepackage{longtable}
\usepackage{array}
\usepackage{amsmath}
\usepackage{siunitx}
\usepackage{amsmath}
\usepackage{booktabs}
\usepackage{rotating}
\usepackage{booktabs}
\usepackage{longtable}
\usepackage{verbatim}
\usepackage[font=small]{caption}

\usepackage[linewidth=1pt]{mdframed}
\allowdisplaybreaks[4]

\providecommand{\U}[1]{\protect\rule{.1in}{.1in}}
\renewcommand{\baselinestretch}{1.2}
\graphicspath{{plots/}}
\newsavebox\CBox

\usepackage{amsthm,thmtools}
\usepackage{mathrsfs}
\usepackage[nopatch=footnote]{microtype}

\usepackage{algorithmicx}%
\usepackage{algpseudocode}%
\usepackage{listings}%
\usepackage{ifpdf}
\usepackage{color}
\usepackage{tikz}
\usetikzlibrary{positioning}
\usepackage{soul}
\newcommand{\Rlogo}{\protect\includegraphics[height=1.8ex,keepaspectratio]{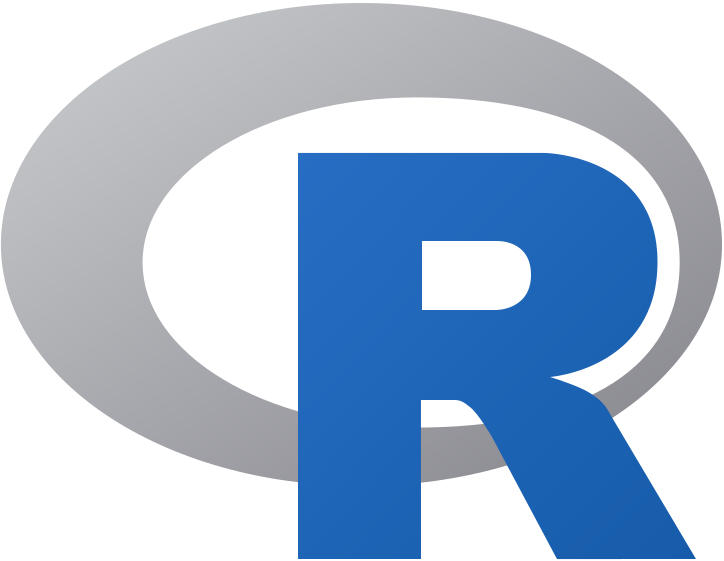}}

\begin{document}
\title{A Dirichlet-Multinomial-Poisson framework for the coherent analysis and forecast of cause-specific mortality}

\def\spacingset#1{\renewcommand{\baselinestretch}
{#1}\small\normalsize} \spacingset{1}

\author{\normalsize Andrea Nigri \orcidlink{0000-0002-2707-3678} \\
\normalsize Department of Social Sciences \\
\normalsize University of Foggia \\
\\
\normalsize Han Lin Shang \orcidlink{0000-0003-1769-6430}  \\
\normalsize Department of Actuarial Studies and Business Analytics \\ 
\normalsize Macquarie University\\
\\
\normalsize Francesco Ungolo \orcidlink{0000-0002-6642-0425}  \\
\normalsize School of Risk and Actuarial Studies \\ 
\normalsize UNSW Sydney
}

\date{}

\maketitle

\centerline{\bf Abstract}
\noindent 

Separate modelling of cause-specific mortality rates and their projections can yield inconsistent forecasts when the sum of deaths by cause does not match the total observed in a population. We develop a hierarchical probabilistic framework for cause-specific mortality counts in which both the total number of deaths and the occurrence of deaths across causes are treated as random. Conditional on the total number of deaths, cause-specific counts follow a multinomial distribution, whereas the total count is modelled using a Poisson distribution, and the vector of cause-of-death probabilities is assigned a Dirichlet distribution. The variation in cause-specific mortality rates by age and calendar year is captured in both the Poisson and Dirichlet models, allowing interpretable demographic patterns while preserving coherence by construction. This model construction naturally preserves the coherence between the sum of deaths by cause and the total mortality. 

The method is exhibited through the analysis of cause-specific mortality rates in the United States and France, sourced from the Human Mortality Database from 1979 to 2023, separately by sex and across ages, with deaths grouped into major cause categories. The empirical analysis uses a rolling 15-year out-of-sample evaluation and compares the proposed model with the standard Lee–Carter model and its compositional extension. The results show that coherent projections can be obtained across countries and sexes, that competitive predictive accuracy is achieved, and that uncertainty is well-calibrated for both total and cause-specific mortality.

\begin{flushleft}

\end{flushleft}

\newpage
\spacingset{1.58}

\section{Introduction}\label{sec1}

Human life expectancy increased considerably in developed countries during the twentieth century due to improved living standards, medical advances, and advances in public health policies. For example, in the United States (U.S.), life expectancy at birth increased from 47.3 years in 1900 to 78.7 years in 2010 \citep{crimmins:2015}. This aspect poses profound epidemiological and social challenges. In addition to the ageing process, the impact of each cause of death changes over time. The proportion of deaths from some causes, such as infectious diseases, is declining dramatically due to medical advances, while others remain persistent or even increase. For public health decision makers, identifying the evolution of mortality by cause is highly relevant for guiding prevention strategies, allocating resources, and designing interventions targeting specific cohorts or generations \citep[see, e.g.,][]{preston2000demography, booth2008, basellini2023thirty}.

Traditional approaches for extrapolating general mortality rates, such as the model of \citeauthor{lee1992modeling}  \citep[see][for a recent review]{basellini2023thirty}, have proven very useful for capturing age-specific trends and projecting overall mortality rates. However, when used for separate analysis of mortality of subpopulations, or by cause of death, these models can produce forecasts that violate basic demographic constraints, such as the addition of cause-specific deaths to total mortality \citep[see, e.g.,][]{wilmoth1995mortality, hyndman2013coherent}. This incoherence yields implausible long-term projections and complicates their interpretation for policy purposes.

From a statistical perspective, modelling mortality by cause is far from trivial. Causes of death are, by definition, mutually exclusive and collectively exhaustive. Their compositional structure implies that independent modelling for each cause, as in the seminal work of \citet{chiang1991competing}, may produce incoherent results, in which the sum of cause-specific forecasts exceeds or falls short of total mortality. This incoherence undermines both actuarial applications and the credibility of long-term public health scenarios, often producing more pessimistic forecasts \citep{wilmoth1995mortality}. A growing research strand has considered the compositional statistical analysis of the proportions of cause-of-death occurrences, embedding statistical frameworks that enforce coherence \citep[see, e.g.,][among others]{oeppen2008coherent, kjaergaard2019forecasting, cardillo2023mortality, cardillo2024tensor}.

A promising development in this direction is the integration of a Dirichlet regression within age-specific mortality modelling. By assuming that the proportion of deaths by cause follows a Dirichlet distribution, it is possible to directly account for the compositional nature of mortality without the need to resort to transformations, as in the work of \cite{ZC17}. Recent contributions show that this approach preserves interpretability while providing a coherent framework for capturing age and temporal variation across causes \citep{grazianinigri}. Furthermore, the Dirichlet framework can be naturally combined with age–period (AP) and their age–period–cohort extensions.

The decomposition of age–period–cohort models unravels the biological effect of ageing, period shocks such as pandemics or medical innovations, and cohort effects reflecting the unique experiences of generations \citep{renshaw2006cohort, hunt2014general}. Extending this paradigm to causes of death enables researchers to uncover hidden regularities in mortality dynamics and improve overall forecasts.

Similar lines of research have focused on age–period models of mortality gaps, particularly when comparing major causes of death such as cancer and circulatory diseases. Using probabilistic models such as the Skellam distribution, these studies capture differences in mortality counts with a good degree of parsimony, highlighting the potential of age–period structures to analyse cause-specific disparities \citep{LLL+19, LanfiutiBaldi2025}. Together, these works highlight a crucial methodological challenge: how to model both the level of total mortality and the composition of deaths by cause while retaining the flexibility to account for age- and period-specific dynamics.

This paper contributes to the literature by proposing a Dirichlet-Multinomial-Poisson (DMP) framework for analysing mortality rates by cause. This framework assumes that death counts by cause follow a multinomial distribution, where both the cause-specific probability of death and the total number of deaths are random and follow Dirichlet and Poisson distributions, respectively. This model induces a double source of dispersion, allowing for the capture of variation in the number of deaths and in the proportions of occurrences across causes. Both types of variation can be explained by covariates such as age and calendar year. The methodology is showcased in the analysis, and the forecast of the mortality rates by cause for the U.S. and France observed between 1979 and 2023. Such rates are analysed separately by sex, using a rolling holdout in which the rates over the last 15 years of the observation period are used to assess the predictive abilities of the two approaches. The results in the DMP framework of this paper are compared with the Lee--Carter (LC) model applied to the analysis of mortality rates by each cause, and with the compositional data analysis (CoDa) method of \cite{kjaergaard2019forecasting}.

The model parameters under the DMP are estimated via a fully Bayesian analysis. This inferential strategy allows us to quantify the uncertainty of the projected rates. This aspect is relevant for policy and planning, as it allows one to assess the model's or parameter's uncertainty levels; compare the estimation or forecast accuracy of different methods more thoroughly; and explore different scenarios based on various assumptions \citep{Chatfield00}.

The predictive uncertainty is obtained by construction from the joint generative model, and it naturally integrates 
\begin{inparaenum}
\item[(i)] stochastic variability in future death counts and cause allocations and 
\item[(ii)] parameter uncertainty through the posterior distribution. 
\end{inparaenum}
This differs from classical approaches such as the LC model, fitted separately by cause, or CoDa methods applied to life-table quantities, where uncertainty bands are often introduced post-hoc, may not propagate all sources of uncertainty in a unified manner, and do not necessarily deliver coherent uncertainty statements simultaneously for totals and causes. By merging demographic reasoning with advanced statistical modelling, our approach simultaneously addresses the epidemiological need to monitor shifting cause-of-death patterns and the statistical demand for coherent and well-calibrated forecasts. This contribution is a step toward a general statistical framework that unifies mortality forecasting and cause-of-death analysis, with applications ranging from longevity risk management to health policy planning.

The remainder of the paper is structured as follows. Section~\ref{sec:pdm} introduces the mortality modelling framework and the DMP framework and its application. Section~\ref{sec:inference} describes the Bayesian inferential framework. In Section~\ref{sec4}, we present a forecasting framework using AP and LC models. Section~\ref{sec:data} details the data used in the empirical analysis. Section~\ref{sec:results} reports and discusses the results, and Section~\ref{sec:conclusion} concludes with final remarks and directions for future research.


\section{Longevity Modelling}\label{sec:pdm}

The statistical modelling of mortality rates captures the variation in three key dimensions: \emph{age}, \emph{period}, and \emph{cause of death}. The age describes the biological and social processes that vary systematically throughout the life cycle; the period reflects contemporaneous shocks and innovations, such as pandemics, wars, or medical advances, that affect all age groups simultaneously throughout a specific calendar year. Finally, the cause of death provides the epidemiological breakdown of mortality into mutually exclusive and collectively exhaustive categories. 

To this end, let $E_{a,t}$ denote the \emph{exposure to risk} (person-years at risk), 
or age group $a$ during period $t$, and let $Y_{a,t}$ denote the random number of deaths among individuals aged $a$ last birthday occurring during period $t$ due to cause $c$. Therefore, the total number of deaths for the age group $a$ and the period $t$, $Y_{a,t}$, can be obtained as the sum of the number of deaths for each cause, $Y_{a,t}=\displaystyle\sum_c Y_{a,t,c}$.

For each age $a$, year $t$, and cause of death $c$, the researcher observes the following:
\begin{itemize}
\item $E_{at}$: the population \emph{exposure to risk}, often approximated by the average between the population aged $a$ at the start of the period $t$ and the population aged $a$ at the end of the same period;  
\item $Y_{atc}$: the number of \emph{deaths} observed at age $a$, year $t$, due to cause $c$.  
\end{itemize}
The corresponding \emph{central death rate} is defined as
\begin{equation}
m_{a,t,c} = \frac{Y_{a,t,c}}{E_{a,t}},
\end{equation}
which represents the cause $c$ average mortality intensity at age $a$ in year $t$. In demographic and actuarial analyses, this quantity is used as an unbiased estimator of the \emph{force of mortality}~$\mu(a,t,c)$, namely the instantaneous risk of death at exact age $a$ and calendar time~$t$ by cause $c$ \citep{preston2000demography}:
\begin{equation}
m_{a,t,c} \approx \mu(x,t,c), \quad x \in [a,a+1).
\end{equation}

For this reason, following the early literature \citep[see, e.g.,][]{brouhns2002poisson}, it is assumed that the random number of deaths $Y_{a,t,c}$ follows a Poisson distribution with mean and variance depending on the central death rate as follows: $$Y_{a,t,c} \sim \text{Poisson}(E_{a,t}\, m_{a,t,c}).$$


This formulation ensures that cause-specific deaths aggregate coherently with the total number of deaths. By the reproducibility property of the Poisson distribution, the sum of the cause-specific death counts is also Poisson-distributed:
\begin{align}\label{eq:poissondeaths}
\displaystyle\sum_{c} Y_{a,t,c}=Y_{a,t} \sim \text{Poisson}\left(E_{a,t}m_{a,t}\right),
\end{align}
where $m_{a,t}=\displaystyle\sum_{c} m_{a,t,c}$.

The proposed DMP framework consists of a hierarchical model, where: 
\begin{inparaenum}
\item[i)] the total number of deaths $Y_{a,t}$ follows a Poisson distribution, and 
\item[ii)] the vector of the number of deaths by each cause $\overline{Y}_{a,t} = \left(Y_{a,t,1},\ldots,Y_{a,t,C} \right)$ follows a Dirichlet-Multinomial (DM) distribution \citep{Mosimann62}. 
\end{inparaenum}
The causes of death are listed for ease of exposition in Table \ref{tab:ICD}. The DM distribution is indexed by two parameters, namely the total number of deaths $Y_{a,t}$ and a vector $\overline{\alpha}_{a,t} = \phi \overline{\gamma}_{a,t}$, where $\overline{\gamma}_{a,t}=\left(\gamma_{a,t,1},\ldots,\gamma_{a,t,C}\right)$, such that $\displaystyle\sum_c \gamma_{a,t,c}=1$, and $\phi$ controls the overdispersion of $\overline{\gamma}_{a,t}$. Thus, $\overline{Y}_{a,t} \sim \text{DM}\left(\overline{\alpha}_{a,t},Y_{a,t} \right)$, with probability mass function
\begin{equation*}
p\left(\overline{y}_{a,t} \mid \overline{\alpha}_{a t}, y_{a,t}\right)=\frac{\Gamma\left(\phi\right) \Gamma\left( y_{a,t}+1\right)}{\Gamma\left(y_{a,t}+\phi\right)} \prod_{c=1}^C \frac{\Gamma\left(y_{a, t, c}+\phi\gamma_{a,t,c}\right)}{\Gamma\left(\phi\gamma_{a,t,c}\right) \Gamma\left(y_{a,t,c}+1 \right)},
\end{equation*}
where $\Gamma(\cdot)$ denotes the Gamma function.

This construction ensures that the cause-specific mortality hazard decomposes as $m_{a,t,c}=m_{a,t}\gamma_{a,t,c}$, which aligns the mean structure of the Dirichlet–Multinomial model with the overall mortality hazard $m_{a,t}$ and the compositional proportions $\gamma_{a,t}$. Each probability $\gamma_{a,t,c}$ is modelled in terms of a softmax function as follows:
\begin{align}
\gamma_{a,t,c} = \frac{\exp\left(\eta_{a,t,c} \right)}{ \displaystyle\sum_c exp\left(\eta_{a,t,c} \right)}.
\end{align}

Where $\eta_{a,t,c}$ is a latent linear predictor for cause $c$ at age $a$ and year $t$.

The DM model combines the Multinomial distribution with Dirichlet-distributed probabilities. In this way, it is possible to account for overdispersion in the vector of cause-specific death counts and to account for the evolution of this composition over the years due to medical and societal advancements. In addition, since the Dirichlet distribution is supported on the simplex, the resulting model ensures compositional coherence and offers a flexible tool for probabilistic modelling and forecasting.

For the data described in Section~\ref{sec:data}, two models are considered for the rate of the Poisson distribution $m_{a,t}$, and for the argument of the softmax function $\eta_{atc}$ to derive the underlying probability of the DM model, namely the AP model in Section~\ref{sbs:ap} and the LC model in Section~\ref{sbs:lc}.

\subsection{Age-Period model}\label{sbs:ap}

This model assumes that age and period are two factors that proportionally affect the rate of the Poisson distribution:
\begin{align}
\log m_{a,t} = \nu_{0} + \delta_a+\pi_{t},
\end{align}
where $\nu_0$ is the baseline $\log$-mortality rate, and $\delta_a$ is the factor deviation corresponding to the effect associated with age $a$, and $\pi_t$ is the proportional effect due to calendar year $t$. Similarly, age and period have an additive effect on the parameter $\eta_{a,t,c}$ through the parameters $\zeta_{a,c}$ and $\lambda_{t,c}$, respectively, for cause $c$ as deviation from the baseline parameter $\varphi_{0,c}$:
\begin{align}
\eta_{a,t,c} = \varphi_{0,c} + \zeta_{a,c}+\lambda_{t,c}.
\end{align}

Mortality rates typically evolve smoothly across ages and over calendar years, and the same pattern is observed in the cause-specific probabilities of death. To reflect this empirical regularity and to avoid overfitting, a smoothing effect is imposed on the parameters $\delta_a$, $\pi_t$, $\eta_a^c$, and $\eta_t^c$ ($c=1,\ldots, C$). This strategy improves the stability of estimates in sparsely populated cells and captures the gradual demographic and epidemiological transitions commonly seen in mortality data. To this end, each effect is modelled as a second-order Gaussian random walk (RW2, \citet{BZ18}):
\begin{align}
\label{eq:smoothdelta}
\Delta^2 \delta_a \equiv \delta_a-2 \delta_{a-1}+\delta_{a-2} & \stackrel{\mathrm{iid}}{\sim} \mathcal{N}\left(0, \sigma_\delta^2\right), \quad a \geq 3 \\
\notag
\Delta^2 \pi_t \equiv \pi_t-2 \pi_{t-1}+\pi_{t-2} & \stackrel{\mathrm{iid}}{\sim} \mathcal{N}\left(0, \sigma_\pi^2\right), \quad t \geq 3\\
\label{eq:smoothzeta}
\Delta^2 \zeta_{a,c} \equiv \zeta_{a,c}-2 \zeta_{a-1,c}+\zeta_{a-2,c} & \stackrel{\mathrm{iid}}{\sim} \mathcal{N}\left(0, \sigma_\zeta^2\right), \quad a \geq 3 \\
\notag
\Delta^2 \lambda_{t,c}\equiv \lambda_{t,c}-2 \lambda_{t-1,c}+\lambda_{t-2,c} & \stackrel{\mathrm{iid}}{\sim} \mathcal{N}\left(0, \sigma_\lambda^2\right), \quad t \geq 3.
\end{align}
For $a=1,2$ and $t=1,2$, age- and period-specific parameters are free parameters to be inferred through the Bayesian inferential procedure outlined in Section~\ref{sec:inference}.

\subsection{Lee-Carter model}\label{sbs:lc}

This model assumes that the $\log$-rate of the Poisson distribution is given by the sum of a baseline mortality level $\nu_0$, an age-specific factor $\delta_a$, and an age-period interaction effect\footnote{This parameterisation is algebraically equivalent to the classical LC model ($a_a=\nu_0+\delta_a,\ b_a=\beta_a$). In our approach, this reparameterisation is useful because it separates the global mortality level ($\nu_0$) from the age profile ($\delta_a$), facilitates centred and smoothed priors on age effects, and improves MCMC stability while preserving the same mean structure. Identification is enforced by $\sum_a \delta_a=0$, $\sum_t \kappa_t=0$, and $\sum_{a} \beta_{a}=1$.}:
\begin{align}
\log m_{a,t} = \nu_0 + \delta_a+\beta_a\kappa_t.
\end{align}

To ensure the identifiability of the model parameters, the classical LC constraints are imposed \citep{hunt2014general}:
\begin{equation*}
\sum_{t=1}^T \kappa_t=0,\qquad \sum_{a=1}^A \beta_a = 1.
\end{equation*}
The age effect $\delta_a$ is smoothed by assuming a RW2 process as in equation \eqref{eq:smoothdelta}, while for the LC model, $\kappa_t$ is assumed to follow a first-order random walk (RW1):
\begin{equation*}
(\kappa_t \mid \kappa_{t-1}) \sim \mathcal{N}(\kappa_{t-1},\sigma_\kappa^2), \qquad t\ge 2.
\end{equation*}

The parameter $\eta_{a,t,c}$ is modelled as an additive function of the age and of a cause-period interaction term. More precisely, such an interaction component uses the same calendar year effect $\kappa_t$ as used in the $\log$-rate of the Poisson model.

This approach allows for the capture of systematic reallocations of the total number of deaths across causes:
\begin{equation*}
\eta_{a,t,c}=\varphi_{0,c}+\zeta_{a,c}+\vartheta_c \kappa_t,
\end{equation*}
where $\zeta_{a,c}$ is an age-by-cause smooth deviation and $\vartheta_c$ is a cause-specific loading to measure the interaction between cause of death probability and period.

The cause-period interaction uses the same calendar-year effect $\kappa_t$ as in the Poisson $\log$-rate. In the LC--DM model, Poisson and Dirichlet--Multinomial parameters are estimated \emph{jointly} in a single Bayesian hierarchy (i.e., no two-stage plug-in for $\kappa_t$), as made explicit by the joint likelihood in Section~\ref{sec:inference}.

The age cause-specific effect $\zeta_{a,c}$ is also smoothed through an RW2 process as in~\eqref{eq:smoothzeta}.

\section{Inference}\label{sec:inference}

\subsection{Prior distribution}\label{sec:3.1}

The statistical analysis of the cause-specific mortality rates in Section~\ref{sec:data} is conducted under non-informative prior assumptions. All parameters are assumed to be pairwise independent \textit{a priori}, so that prior information does not impose artificial dependencies across the components of the model.

Hence, diffuse priors are assigned to the intercepts and level parameters governing total mortality and the cause-specific compositions. The prior distribution for each model parameter under the AP and LC specifications is summarised in Table~\ref{tab:priors}. 

\begin{center}
\tabcolsep 0.5in
\begin{longtable}{@{}lll@{}}
\caption{Prior specifications and identifiability constraints for the AP--DM and LC--DM models.}\label{tab:priors}\\
\toprule
Model block & Parameter(s) & Prior / Constraint \\ 
\midrule
\endfirsthead

\toprule
Model block & Parameter(s) & Prior / Constraint \\ 
\midrule
\endhead

\midrule
\multicolumn{3}{r}{Continued on next page} \\
\endfoot
\endlastfoot
\multicolumn{3}{@{}l}{\textbf{AP--DM: total mortality (Poisson)}}\\
& $\nu_0$ & $\mathcal{N}(0,5^2)$ \\
& $\delta_1,\delta_2$ & $\mathcal{N}(0,5^2)$ \\
& $\pi_1,\pi_2$ & $\mathcal{N}(0,1)$ \\
& $\sigma_\delta$ & $\mathcal{N}^+(0,1^2)$ \\
& $\sigma_\pi$ & $\mathcal{N}^+(0,0.3^2)$ \\
\addlinespace
\multicolumn{3}{@{}l}{\textbf{AP--DM: composition (Dirichlet--Multinomial), for each cause $c=1,\ldots,C$}}\\
& $\varphi_{0,c}$ & $\mathcal{N}(0,5^2)$ \\
& $\zeta_{1,c},\zeta_{2,c}$ & $\mathcal{N}(0,5^2)$ \\
& $\lambda_{1,c},\lambda_{2,c}$ & $\mathcal{N}(0,1)$ \\
& $\sigma_\zeta$ & $\mathcal{N}^+(0,1^2)$ \\
& $\sigma_\lambda$ & $\mathcal{N}^+(0,0.3^2)$ \\
\addlinespace
\multicolumn{3}{@{}l}{\textbf{AP--DM: overdispersion}}\\
& $\phi$ & $\log\mathcal{N}(\phi_\mu,\phi_\sigma)$ \\
\addlinespace
\multicolumn{3}{@{}l}{\textbf{AP--DM: identifiability constraints}}\\
&  &
$\sum_{a=1}^A \delta_a = 0,\ \ \sum_{t=1}^T \pi_t = 0,\ \ \sum_{c=1}^C \varphi_{0,c}=0,$\\
& &
$\sum_{a=1}^A \zeta_{a,c}=0,\ \ \sum_{t=1}^T \lambda_{t,c}=0\ \ (\forall c).$\\
\midrule
\multicolumn{3}{@{}l}{\textbf{LC--DM: total mortality (LC + Poisson)}}\\
& $\nu_0$ & $\mathcal{N}(0,10^2)$ \\
& $\delta_1,\delta_2$ & $\mathcal{N}(0,5^2)$ \\
& $\beta_a$ & $\mathcal{N}^+(0,1^2)$ \\
& $\kappa_1$ & $\mathcal{N}(0,1^2)$ \\
& $\sigma_\delta$ & $\mathcal{N}^+(0,0.5^2)$ \\
& $\sigma_\kappa$ & $\mathcal{N}^+(0,0.5^2)$ \\
\addlinespace
\multicolumn{3}{@{}l}{\textbf{LC--DM: composition (Dirichlet--Multinomial), for each cause $c=1,\ldots,C$}}\\
& $\varphi_{0,c}$ & $\mathcal{N}(0,1^2)$ \\
& $\zeta_{1,c},\zeta_{2,c}$ & $\mathcal{N}(0,1^2)$ \\
& $\vartheta_c$ & $\mathcal{N}(0,0.5^2)$ \\
& $\sigma_\zeta$ & $\mathcal{N}^+(0,0.4^2)$ \\
\addlinespace
\multicolumn{3}{@{}l}{\textbf{LC--DM: overdispersion}}\\
& $\phi$ & $\log\mathcal{N}(\phi_\mu,\phi_\sigma)$ \\
\addlinespace
\multicolumn{3}{@{}l}{\textbf{LC--DM: identifiability constraints}}\\
&  &
$\sum_{t=1}^T \kappa_t = 0,\ \ \sum_{a=1}^A \beta_a = 1,\ \ \sum_{a=1}^A \delta_a=0,$\\
& &
$\sum_{c=1}^C \varphi_{0,c}=0,\ \ \sum_{a=1}^A \zeta_{a,c}=0\ \ (\forall c).$\\
\bottomrule
\end{longtable}
\end{center}

The model includes a small set of parameters (e.g., intercepts, dispersion, and cause-specific loadings), while the remaining parameters are indexed by age and period (e.g., $\delta_a$, $\beta_a$, $\kappa_t$, $\zeta_{a,c}$). Therefore, the total dimension depends on the data and increases with the number of ages ($A$) and periods ($T$) considered.

These choices aim to minimise the influence of prior specification on posterior inference, allowing the data to fully determine the posterior distribution of the parameters.

\subsection{Likelihood}\label{sec:3.2}

Let $\theta$ denote the full vector of model parameters indexing the DMP under the assumptions of the AP or LC model. For each age group $a$ and calendar year $t$, the observed death counts are conditionally independent given the corresponding parameters.

The likelihood function for $\theta$, conditional on exposures $E_{a,t}$ and observed death counts by cause $\mathbf{Y}$, factorises into a Poisson component for total deaths and a DM component for the cause composition:
\begin{equation}
\mathcal{L}(\boldsymbol{\theta} ; \mathbf{Y}, \mathbf{E})=\prod_{a=1}^A \prod_{t=1}^T \underbrace{p\left(y_{a,t} \mid E_{a,t} m_{a,t}\right)}_{\text {Poisson}\left(E_{a,t} m_{a,t}\right)} \times \underbrace{p\left(\overline{y}_{a,t} \mid \phi \overline{\gamma}_{a,t}, y_{a,t}\right)}_{\text {DM}\left(\phi \overline{\gamma}_{a,t}, y_{a,t}\right)},
\end{equation}
where $\overline{y}_{a,t}=(y_{a,t,1} \cdots,y_{a,t,C}$), $y_{a,t}=\displaystyle\sum_{c=1}^C y_{a,t,c}$, and  $\boldsymbol{\theta}$ can be $\boldsymbol{\theta}_{\mathrm{AP}}=
\{\nu_0,\delta,\pi,\phi,\zeta,\lambda,
\sigma_\delta,\sigma_\pi,\sigma_\zeta,\sigma_\lambda\},$ $\boldsymbol{\theta}_{\mathrm{LC}}=
\{\nu_0,\delta,\beta,\kappa,\phi,\zeta,\vartheta,
\sigma_\delta,\sigma_\kappa,\sigma_\zeta\}$.

The Poisson component governs the uncertainty around the total number of deaths, while the Dirichlet--Multinomial component captures the allocation of these deaths across causes and accommodates overdispersion and compositional structure. 

\subsection{Posterior inference}\label{sbs:postinfhmc}

Posterior inference follows from the product of the likelihood in Section~\ref{sec:3.2} and the prior specification densities, as outlined in Section~\ref{sec:3.1}. The resulting posterior distribution does not admit a closed-form expression.

The posterior distribution of the parameters is estimated using the Hamiltonian Monte Carlo (HMC) algorithm, a class of Metropolis-Hastings samplers that enables faster, more effective exploration of the parameter space. This is implemented in Stan \citep{stan} through the \Rlogo\ interface using the \Rlogo\ package \texttt{rstan}. The sampler is automatically tuned through the ``no U-turns sampler" of \citet{hoffman2014no}. The key feature of the HMC sampler is the ability to generate proposals that follow informed trajectories through the parameter space, taking advantage of the geometry of the target density. This results in more efficient exploration, particularly in high-dimensional parameter spaces with highly correlated parameters, thereby reducing autocorrelation and accelerating convergence.

\section{Projection}\label{sec4}

Let $H$ denote the forecast horizon and $\mathcal{T}^{+}=\{T+1,\ldots,T+H\}$ the set of future periods. The primary goal is to obtain coherent projections of age-specific mortality rates by cause such that the sum of cause-specific rates equals the corresponding total rate at each age and period. This ensures epidemiological interpretability and avoids incoherence between aggregate and disaggregate forecasts.

If official population projections are available, exposures $E_{a,t}$ for $t \in \mathcal{T}^{+}$ can be incorporated to produce forecasts of future death counts. However, the fundamental outputs of our model are forecasts on the \emph{rate scale}: the total death rate $m_{a,t}$ and the vector of cause-specific shares $\gamma_{a,t,c}$. This strategy guarantees that the forecasts remain internally consistent across causes, even when exposures are unknown or uncertain.

\paragraph{Forecasting under AP--DM.} Future-period effects can be obtained by leveraging their assumed RW2 dynamics beyond $T$. For each posterior draw and $t \in \mathcal{T}^{+}$,
\begin{align*}
\pi_t &= 2\pi_{t-1} - \pi_{t-2} + \varepsilon_t^{(\pi)}, \quad \varepsilon_t^{(\pi)} \sim \mathcal{N}(0,\sigma_\pi^2),\\
\lambda_{t,c} &= 2\lambda_{t-1,c} - \lambda_{t-2,c} + \varepsilon_{t,c}^{(\lambda)}, \quad \varepsilon_{t,c}^{(\lambda)} \sim \mathcal{N}(0,\sigma_{\lambda}^2), \quad c=1,\ldots,C.
\end{align*}
The age effects $\delta_a$ and $\zeta_{a,c}$ remain fixed, with centring constraints applied for numerical stability.

Therefore, the projected total rate $m_{a,t}$ and the corresponding cause shares $\gamma_{a,t,c}$ are computed as follows: 
\begin{equation*}
m_{a,t} = \exp\{\nu_0+\delta_a+\pi_t\}, \qquad
\gamma_{a,t,c} = \frac{\exp\{\varphi_{0,c}+\zeta_{a,c}+\lambda_{t,c}\}}{\sum_{c'}\exp\{\varphi_{0,c'}+\zeta_{a,c'}+\lambda_{t,c'}\}}, \qquad \sum_c \gamma_{a,t,c}=1.
\end{equation*}

\paragraph{Forecasting under LC--DM.} For LC--DM we simulate the future $\kappa_t$ by extending the RW1 dynamics:
\begin{equation*}
\kappa_t = \kappa_{t-1} + \varepsilon_t^{(\kappa)},\qquad \varepsilon_t^{(\kappa)}\sim \mathcal{N}(0,\sigma_\kappa^2),
\qquad t\in\mathcal{T}^{+}.
\end{equation*}
The age effects $\delta_a$ and $\zeta_{ac}$ remain fixed (with centering constraints), while $\beta_a$ and $\vartheta_c$ are time-invariant loadings.

For LC--DM, the corresponding projected quantities are
\begin{equation*}
m_{at} = \exp\{\nu_0+\delta_a+\beta_a\kappa_t\}, \qquad
\gamma_{atc} = \frac{\exp\{\varphi_{0,c}+\zeta_{a,c}+\vartheta_c\kappa_t\}}{\sum_{c'}\exp\{\varphi_{0,c'}+\zeta_{a,c'}+\vartheta_{c'}\kappa_t\}}, \qquad \sum_c \gamma_{a,t,c}=1.
\end{equation*}

For both AP and LC models, the combination of these projected quantities directly yields cause-specific mortality rates $m_{a,t,c} = m_{a,t}\,\gamma_{a,t,c}$, which are coherent by construction, since $\sum_c m_{atc} = m_{at}$ for all $(a,t)$. Thus, forecasts naturally provide both the aggregate dynamics and their decomposition across causes \footnote{If exposures $E_{at}$ are available, predictive counts can be simulated via
\begin{equation*}
Y_{a,t}\sim \text{Poisson}(E_{a,t} m_{a,t}), \qquad 
\overline{Y}_{a,t}\sim \text{DM}(\phi \overline{\gamma}_{a,t}),
\end{equation*}
with $\sum_c Y_{a,t,c}=Y_{a,t}$ by design.}.

\section{Data}\label{sec:data}

The models outlined in Section~\ref{sec:pdm} are showcased through an analysis of mortality rates, as available in \cite{HMD}. These are extracted from the period life tables by sex and 5-year age group (0--100+) of the HMD from 1979 to 2023. Death counts by cause and exposure are sourced from \citet{HMCD} (HCD) for both gender in the U.S. and France, and the proportions of deaths by cause, age, and calendar year are computed. The HCD Database provides high-quality cause-specific mortality data in five-year age groups. The taxonomy of the causes of death is given by the International Classification of Diseases (ICD) of the World Health Organisation and consists of three aggregation levels: full list, intermediate list, and short list. Each classification has been developed using the same criteria across all countries, ensuring its homogeneity and comparability. Using these data, one can develop a universal and standardised methodology to redistribute deaths across 104 disease categories within five-year age groups, avoiding issues due to ICD revisions and ensuring cross-country comparability despite differing coding practices. Starting with the HCD classification, the main causes of death are grouped by age up to 100+.

To understand the conditions that may have affected the life expectancy of the U.S. population, six major causes have been considered: Infectious, neoplasms, cardiovascular diseases (CVDs), respiratory diseases, external diseases, and `other', which includes all remaining causes. Table~\ref{tab:ICD} summarises the information on cause-groups by the ICD10 classification system.
\begin{table}[!htb]
\caption{ICD codes and classification into six broad categories.}\label{tab:ICD}
\centering
\rotatebox{0}{
\begin{tabular}{@{}l|l|l@{}}
\toprule
\textbf{Title}              & \textbf{ICD 10}       & \textbf{Cause}     \\
\midrule
Infectious                   & A00--B99              & Infectious          \\ 
Neoplasms                   & C00--D48              & Neoplasms          \\ 
Heart diseases              & I00--I52              & CVD                \\
Cerebrovascular diseases    & G45, I60--I69         & CVD                \\ 
Other and unspecified disorders of the circulatory system & I70--I99 & CVD \\
Acute respiratory diseases  & J00--J22, U04         & Respiratory \\
Other respiratory diseases  & J30--J98              & Respiratory \\ 
External causes             & V01--Y98              & External \\ 
All other causes            & Remaining ICD codes   & Other    \\ 
\bottomrule
\end{tabular}
}
\end{table}

Data from 1989–2008 (1989–2007 for France) are used to estimate the posterior distribution of the model parameters, while the remaining data (2009–2023 for the U.S. and 2008–2022 for France) are used as held-out set to assess the predictive performance of the proposed approach, using a 15-year window for both countries. In a second modelling exercise, whose results are reported in Appendix~\ref{Appendix:A}, the data from 1979 until 1988 are also used to fit the model, and the predictive performance is again assessed over the period 2009–2023 for the U.S. and 2008–2022 for France.

The HMC sampler described in Section~\ref{sbs:postinfhmc} was run three times in parallel, yielding three chains each with $3,000$ samples, where the first 1,500 draws were used as a warm-up. The convergence of the posterior distribution of the parameters was assessed through the $\widehat{R}$ diagnostic in Tables~\ref{tab:fr_posteriors_scalars_all} and~\ref{tab:us_posteriors_scalars_all}, \citep[see, e.g.,][]{gelman1992inference, stan}.

\section{Results}\label{sec:results}

First of all, for most parameters the $\widehat{R}$ is below the suggested 1.05 threshold, indicating that the three chains converged towards a stationary posterior distribution.

Posterior summaries for the scalar parameter of the Bayesian AP and LC models, by sex and in-sample year in both countries under analysis, are presented in Appendix~\ref{App:C}.

Let $\log\widehat m_{a,t,c}$ denote the predicted value of $\log m_{a,t,c}$, namely the death rate in log scale for age $a$, year $t$, cause $c$ under the  model considered. The accuracy over the evaluation set $\mathcal{I}$ (all ($a,t,c$) in the out-of-sample years) is assessed by calculating the mean absolute error (MAE) and the root mean squared error (RMSE):
\begin{align*}
\mathrm{MAE} &= \frac{1}{|\mathcal{I}|}\sum_{(a,t,c)\in\mathcal{I}}\bigl|\log m_{a,t,c}-\log \widehat m_{a,t,c}\bigr|,\\
\mathrm{RMSE} &= \Biggl[\frac{1}{|\mathcal{I}|}\sum_{(a,t,c)\in\mathcal{I}}\bigl(\log m_{a,t,c}-\log \widehat m_{a,t,c}\bigr)^{2}\Biggr]^{1/2}.
\end{align*}

Both measures are calculated also separately by cause, together with their correspondent in-sample summaries, and then compared with those obtainable under the LC model fitted by each cause singly, and with the CoDA approach. The key features of these two competing approaches are summarised in Appendix~\ref{app:benchmarks_lc_coda}.

\subsection{United States (U.S.)}\label{sec:6.2}

The DMP-AP, DMP-LC, CoDa, and LC are compared under a rolling 15-year out-of-sample evaluation. Table~\ref{tab:us_total_train_oos_by_sex} shows the in-sample (Train) and the out-of-sample (OOS) MAE and RMSE by sex for the total number of deaths. 

\paragraph{Total mortality} Table~\ref{tab:us_total_train_oos_by_sex} shows a clear advantage for the DMP-LC specification on the total mortality surface. For both females and males, DMP-LC achieves the lowest in-sample and out-of-sample errors. This is consistent with the role of the latent period index in capturing the dominant secular trend while borrowing strength across ages through the smooth age structure. DMP-AP typically ranks next among the coherent Bayesian specifications, indicating that an additive age--period structure remains competitive. The LC factorisation yields a sharper representation of the time trend in this application.
\begin{table}[!htb]
\centering
\caption{U.S. -- Total mortality performance (Train vs OOS) by sex. Lower is better. Best in \textbf{bold}, second best \underline{underlined} (within each sex).}
\label{tab:us_total_train_oos_by_sex}
\small
\setlength{\tabcolsep}{19pt}
\renewcommand{\arraystretch}{1.15}
\begin{tabular}{@{}llrrrr@{}}
\toprule
\textbf{Sex} & \textbf{Model} & \textbf{Train RMSE} & \textbf{Train MAE} & \textbf{OOS RMSE} & \textbf{OOS MAE}\\
\midrule
Females & DMP-AP & \underline{0.0084} & \underline{0.0027} & 0.0208 & 0.0072\\
        & DMP-LC & \textbf{0.0047} & \textbf{0.0015} & \textbf{0.0090} & \textbf{0.0039}\\
        & CoDa      & 0.1475 & 0.0833 & 0.1464 & 0.0808\\
        & LC        & 0.0101 & 0.0034 & \underline{0.0134} & \underline{0.0047}\\
\midrule
Males   & DMP-AP & 0.0174 & 0.0058 & 0.0199 & 0.0074\\
        & DMP-LC & \textbf{0.0038} & \textbf{0.0015} & \textbf{0.0110} & \textbf{0.0051}\\
        & CoDa      & 0.1690 & 0.1011 & 0.1663 & 0.0944\\
        & LC        & \underline{0.0158} & \underline{0.0055} & \underline{0.0179} & \underline{0.0070}\\
\bottomrule
\end{tabular}
\end{table}

By contrast, the LC benchmark (fitted independently in each series in our implementation) is consistently less accurate than DMP-LC on totals, despite being the second-best method overall on aggregate metrics in several cells. This suggests that the Bayesian formulation, with explicit smoothing priors and joint inference, offers a real gain over the classical LC baseline. The CoDa approach exhibits substantially larger errors in total mortality under this evaluation, which is unsurprising given that CoDa methods are designed to model death distributions in a life-table (simplex) geometry rather than directly modelling death rates.

\paragraph{Mortality by cause}

Table~\ref{tab:us_oos_by_cause_model_sexcols} shows the out-of sample RMSE and MAE by sex and cause. The results are heterogeneous across causes, but two empirical patterns are notable. The DMP-AP and DMP-LC models are systematically among the top performers (best or second-best) for several major causes, and they do so while preserving internal coherence between totals and causes. In particular, DMP-LC is competitive for neoplasms (NEOP) in both sexes and performs strongly for cardiovascular diseases (CVD) in males on RMSE. At the same time, DMP-AP remains competitive for CVD, looking at the MAE and for external causes (EXT) in females. These findings are consistent with the intention of the modelling. Indeed, DMP-LC uses a single latent time factor to drive both the overall mortality level and systematic reallocations across causes. In contrast, DMP-AP allows for a more flexible (additive) period component in both the total and the composition. 

In addition, the cause-specific LC and the CoDa can perform better when analysing specific causes, yet these gains do not translate uniformly across the full system. This is an important distinction: specifically, when the LC is fitted independently by cause (a common practice), it does not guarantee coherence between the sum of cause-specific forecasts and the forecast of total mortality, potentially producing internally inconsistent projections. \citet{kjaergaard2019forecasting} discuss this general limitation of independently forecasting cause-specific rates and motivate coherent alternatives that respect the competing-risk structure. In a similar manner, CoDa methods enforce coherence in the compositional (life-table deaths) space by construction, but they operate on transformed life-table quantities (e.g., centred log-ratio (CLR)-transformed life-table deaths) and require an additional layer of analysis; these features can make performance sensitive to the particular evaluation target (rates vs. life-table deaths) and to data idiosyncrasies, especially for smaller causes.
\setlength{\tabcolsep}{28pt}
\renewcommand{\arraystretch}{0.88}
\begin{longtable}{@{}llrrrr@{}}
\caption{U.S. -- Out-of-sample performance by cause and model, with sex shown as column groups. Lower is better for RMSE/MAE. Best in \textbf{bold}, second best \underline{underlined} (within each cause $\times$ sex, separately for RMSE and MAE).}\label{tab:us_oos_by_cause_model_sexcols}\\
\toprule
 &  & \multicolumn{2}{c}{\textbf{Females}} & \multicolumn{2}{c}{\textbf{Males}}\\
\cmidrule(lr){3-4}\cmidrule(lr){5-6}
\textbf{Cause} & \textbf{Model} & \textbf{RMSE} & \textbf{MAE} & \textbf{RMSE} & \textbf{MAE}\\
\midrule
\endfirsthead
\toprule
 &  & \multicolumn{2}{c}{\textbf{Females}} & \multicolumn{2}{c}{\textbf{Males}}\\
\cmidrule(lr){3-4}\cmidrule(lr){5-6}
\textbf{Cause} & \textbf{Model} & \textbf{RMSE} & \textbf{MAE} & \textbf{RMSE} & \textbf{MAE}\\
\midrule
\endhead
\midrule
\multicolumn{6}{r}{\small continued on next page}\\
\endfoot
\bottomrule
\endlastfoot

CVD & DMP-AP & \textbf{0.0060} & \underline{0.0024} & 0.0132 & \textbf{0.0042}\\
& DMP-LC & 0.0134 & 0.0056 & \textbf{0.0112} & 0.0052\\
& CoDa      & \underline{0.0066} & \textbf{0.0020} & 0.0146 & \underline{0.0050}\\
& LC        & 0.0158 & 0.0068 & \underline{0.0124} & 0.0061\\
\addlinespace

EXT & DMP-AP & \textbf{0.0004} & \textbf{0.0002} & 0.0008 & 0.0004\\
& DMP-LC & \underline{0.0004} & \underline{0.0002} & \textbf{0.0004} & \underline{0.0003}\\
& CoDa      & 0.0004 & 0.0002 & 0.0007 & 0.0004\\
& LC        & 0.0017 & 0.0007 & \underline{0.0005} & \textbf{0.0003}\\
\addlinespace

INF & DMP-AP & 0.0006 & 0.0002 & 0.0012 & 0.0006\\
& DMP-LC & 0.0005 & 0.0002 & \underline{0.0008} & \underline{0.0004}\\
& CoDa      & \textbf{0.0002} & \textbf{0.0001} & 0.0012 & 0.0005\\
& LC        & \underline{0.0004} & \underline{0.0002} & \textbf{0.0004} & \textbf{0.0002}\\
\addlinespace

NEOP & DMP-AP & 0.0024 & 0.0011 & 0.0030 & 0.0013\\
& DMP-LC & \textbf{0.0007} & \textbf{0.0004} & \underline{0.0023} & \underline{0.0013}\\
& CoDa      & 0.0011 & \underline{0.0005} & \textbf{0.0017} & \textbf{0.0007}\\
& LC        & \underline{0.0008} & 0.0005 & 0.0023 & 0.0014\\
\addlinespace

OTHER & DMP-AP & \underline{0.0081} & \underline{0.0032} & \underline{0.0080} & \textbf{0.0029}\\
& DMP-LC & 0.0100 & 0.0035 & \textbf{0.0076} & \underline{0.0030}\\
& CoDa      & \textbf{0.0069} & \textbf{0.0023} & 0.0095 & 0.0036\\
& LC        & 0.0279 & 0.0105 & 0.0310 & 0.0127\\
\addlinespace

RESP & DMP-AP & 0.0080 & 0.0026 & 0.0058 & 0.0021\\
& DMP-LC & 0.0044 & 0.0016 & 0.0056 & 0.0022\\
& CoDa      & \underline{0.0035} & \textbf{0.0010} & \textbf{0.0044} & \textbf{0.0017}\\
& LC        & \textbf{0.0033} & \underline{0.0012} & \underline{0.0046} & \underline{0.0019}\\
\end{longtable}

Table~\ref{tab:us_oos_mean_across_causes_by_sex} summarises arithmetic averages of RMSE and MAE by causes. For males, the RMSE is lowest for the DMP-LC model, while the MAE is lowest for the DMP-AP model. These results indicate that the coherent Bayesian approaches offer the best overall trade-off when aggregating the two measures of predictive performance across the causes. For females, the CoDa model yields the lowest mean errors across causes. However, this result should be treated with caution. In the overall analysis, the same method performs poorly for all-cause mortality under the current rate-based metric. This suggests that the compositional life-table model is optimising something different from what is being evaluated here.
\begin{table}[!htb]
\centering
\caption{U.S. -- OOS mean performance across causes by sex. Lower is better. Best in \textbf{bold}, second best \underline{underlined} (within each sex).}
\label{tab:us_oos_mean_across_causes_by_sex}
\setlength{\tabcolsep}{48pt}
\renewcommand{\arraystretch}{1.15}
\begin{tabular}{@{}llrr@{}}
\toprule
\textbf{Sex} & \textbf{Model} & \textbf{RMSE mean} & \textbf{MAE mean}\\
\midrule
Female  & DMP-AP & \underline{0.0042}    & \underline{0.0016}\\
        & DMP-LC & 0.0049                & 0.0019\\
        & CoDa      & \textbf{0.0031}       & \textbf{0.0010}\\
        & LC        & 0.0083                & 0.0033\\
\midrule
Male    & DMP-AP & \underline{0.0053}    & \textbf{0.0019}\\
        & DMP-LC & \textbf{0.0047}       & 0.0021\\
        & CoDa      & 0.0054                & \underline{0.0020}\\
        & LC        & 0.0085                & 0.0038\\
\bottomrule
\end{tabular}
\end{table}

In contrast, DMP-LC produces the most accurate total mortality forecasts while remaining competitive across multiple causes, which is precisely the objective of coherent forecasting in applications where both the level and the decomposition of mortality rates are of relevance.

These results support two conclusions. 
\begin{inparaenum}[1)]
\item The Bayesian LC--DM formulation provides the most accurate forecasts for total mortality in both sexes over the forecast horizon, while maintaining coherence between total and cause-specific rates by construction. 
\item Although some benchmarks can outperform our models for specific causes, they do not deliver the same combination of 
\begin{inparaenum}
\item[(i)] accurate totals, 
\item[(ii)] coherent decomposition, and 
\item[(iii)] a single joint probabilistic model that yields predictive uncertainty for both the level and the composition. 
\end{inparaenum}
\end{inparaenum}
This joint/coherent perspective is essential for public-health planning, because internally inconsistent cause forecasts can lead to implausible decompositions even when individual cause trajectories appear reasonable \citep{kjaergaard2019forecasting}.
\begin{figure}[!htb]
\centering
\includegraphics[width=1\linewidth]{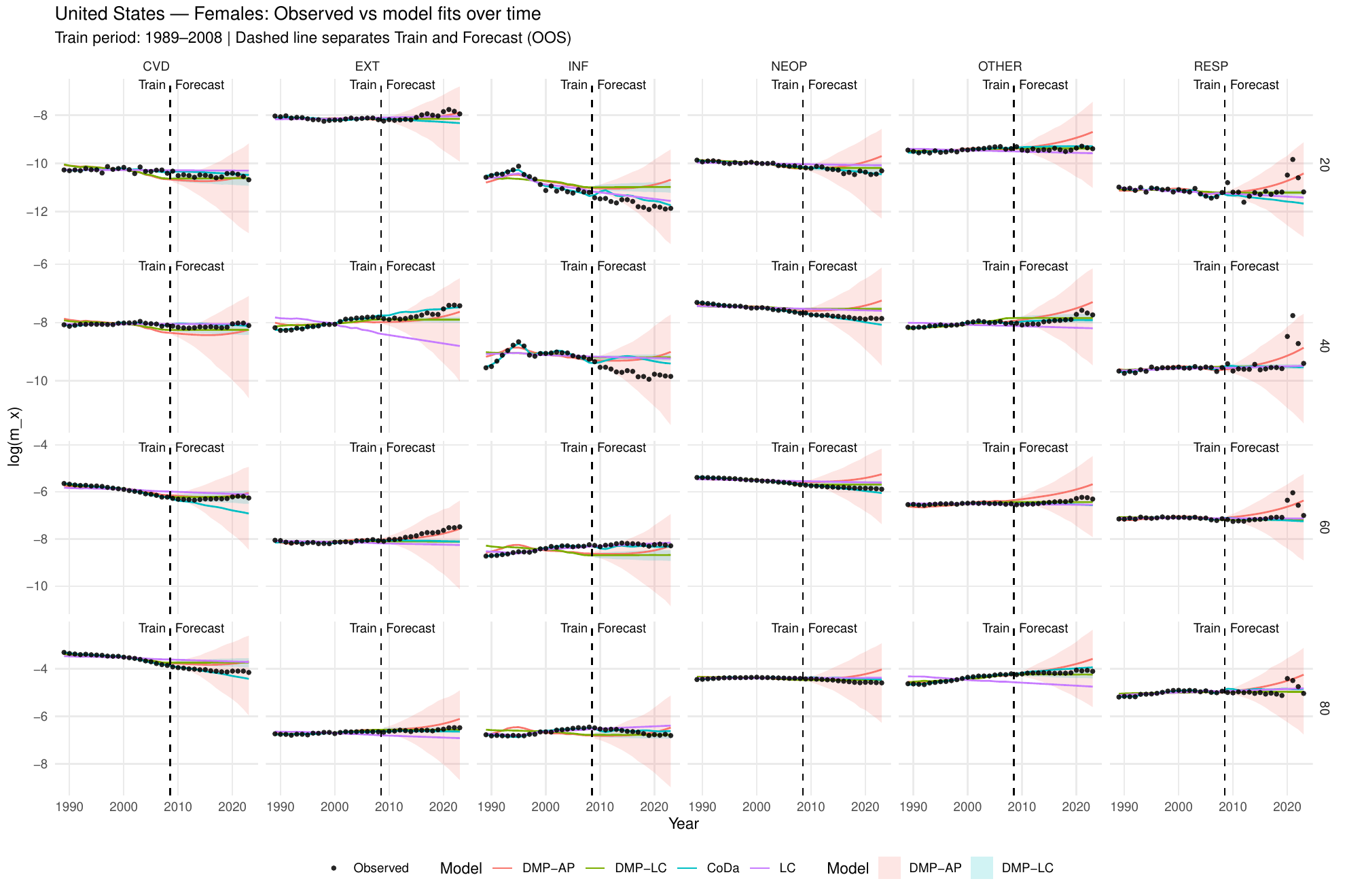}
\caption{\small U.S. Female: Observed (black points) and fitted/forecast $\log$ mortality rates by cause for selected ages (20, 40, 60, 80). The vertical dashed line marks the end of the in-sample period (1989--2008) and the start of the out-of-sample forecasts.}\label{fig:Female}
\end{figure}

\begin{figure}[!htb]
\centering
\includegraphics[width=1\linewidth]{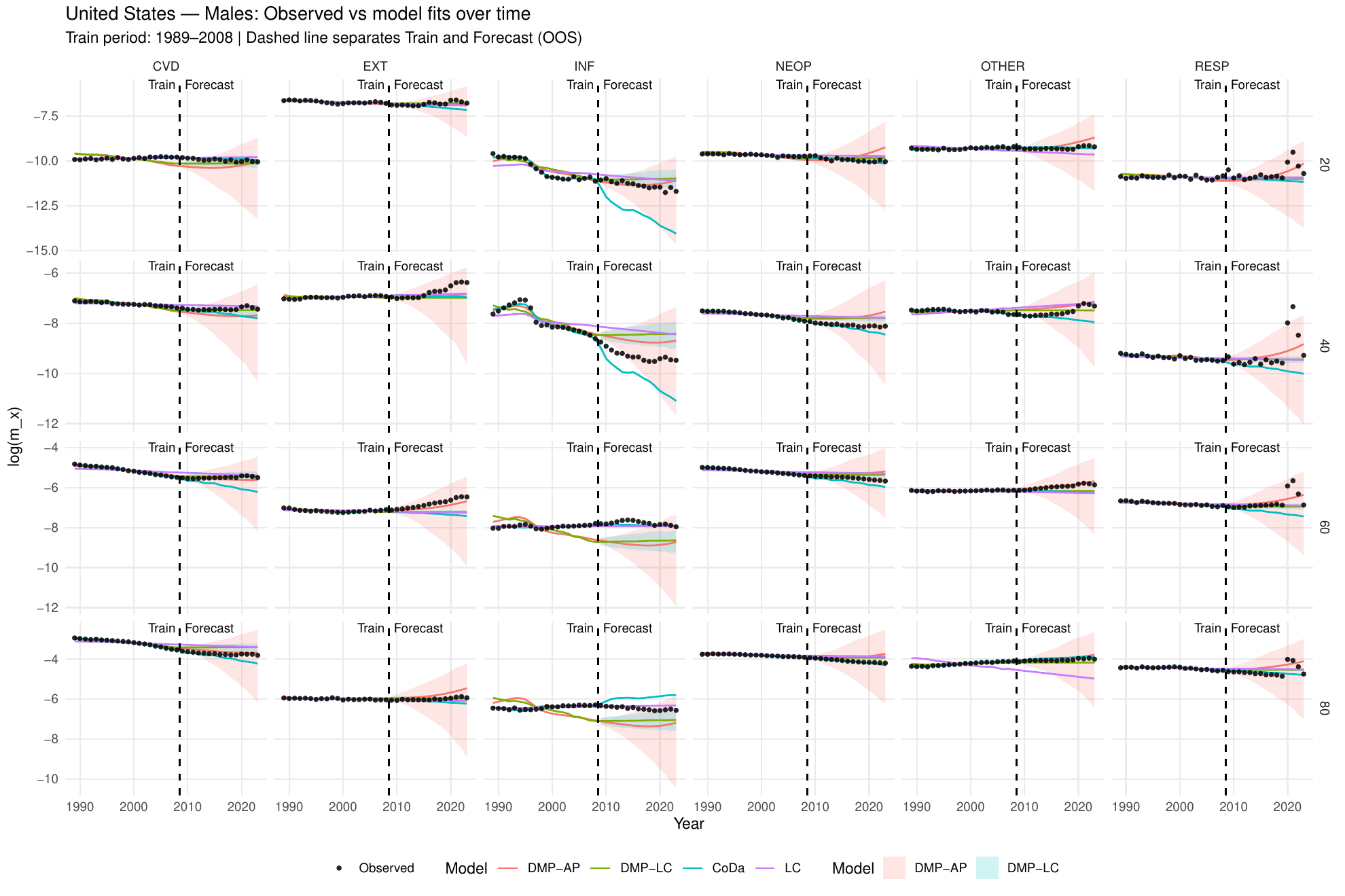}
\caption{U.S. Male: Observed (black points) and fitted/forecast $\log$ mortality rates by cause for selected ages (20, 40, 60, 80). The vertical dashed line marks the end of the in-sample period (1989--2008) and the start of the out-of-sample forecasts.}\label{fig:Male}
\end{figure}

Figures~\ref{fig:Female} and~\ref{fig:Male} display the observed $\log$-mortality rates by cause over time for selected ages (20, 40, 60, 80), with the vertical dashed line marking the transition from the in-sample window (1989--2008) to the forecasting window. The observed rates (black dots) are generally fitted well by the two coherent DMP specifications (DMP-AP and DMP-LC). Their predictive intervals widen after 2008, reflecting the increasing uncertainty over the forecast horizon. This behaviour is particularly evident in causes with higher variability and/or weaker signals at young ages, where credible intervals expand substantially in OOS.

Across both sexes, DMP-LC tends to provide the most stable extrapolations in panels dominated by smooth trends (e.g., CVD and NEOP at older ages), while DMP-AP remains competitive when the post-2008 dynamics resemble a continuation of the pre-2008 additive period pattern. In contrast, the non-coherent benchmarks can display more pronounced panel-specific departures in OOS. The single-cause LC fits sometimes show systematic drift relative to the observed points in certain age--cause combinations, and CoDa exhibits occasional strong deviations in some infectious/respiratory panels (notably at younger ages), consistent with sensitivity to the life-table/compositional representation when evaluated on rate trajectories. 

Overall, the plots support the previous findings: coherent Bayesian models provide excellent in-sample calibration, plausible and more competitive out-of-sample trajectories, and uncertainty bands that expand appropriately beyond the in-sample boundary, while maintaining internally consistent cause decompositions at each age and period.

\subsection{France}

We report results for France using the same evaluation design as for the U.S.: the last 15 calendar years are held out as out-of-sample, and the preceding years form the in-sample set. Performance is evaluated on 
\begin{inparaenum}
\item[(i)] the total mortality surface and 
\item[(ii)] cause-specific mortality, using RMSE and MAE (lower is better), reported separately by sex in Tables~\ref{tab:fr_total_train_oos_by_sex_1989}--\ref{tab:fr_oos_mean_across_causes_by_sex_1989}.
\end{inparaenum}

Table~\ref{tab:fr_total_train_oos_by_sex_1989} shows that the DMP-LC specification dominates total mortality forecasting for both sexes in France when in-sample starts in 1989. For females, DMP-LC achieves the lowest in-sample and out-of-sample errors, with a substantial gap over the alternatives. A striking feature is the very large OOS error in DMP-AP for females on totals, indicating an unstable extrapolation of the level component over the forecast horizon. For males, DMP-LC is again the most accurate method regarding totals, followed by the classical LC benchmark, while CoDa remains clearly less competitive on the rate scale in this aggregate evaluation.
\begin{table}[!htb]
\centering
\caption{France -- Total mortality performance (Train vs OOS) by sex, train from 1989. Lower is better. Best in \textbf{bold}, second best \underline{underlined} (within each sex, separately for each metric).}
\label{tab:fr_total_train_oos_by_sex_1989}
\setlength{\tabcolsep}{16pt}
\renewcommand{\arraystretch}{1.15}
\begin{tabular}{@{}llrrrr@{}}
\toprule
\textbf{Sex} & \textbf{Model} & \textbf{Train RMSE} & \textbf{Train MAE} & \textbf{OOS RMSE} & \textbf{OOS MAE}\\
\midrule
Females & DMP-AP & 0.0092 & 0.0027 & 0.2826 & 0.0647\\
        & DMP-LC & \textbf{0.0036} & \textbf{0.0010} & \textbf{0.0074} & \textbf{0.0028}\\
        & CoDa      & 0.1527 & 0.0838 & 0.1413 & 0.0738\\
        & LC        & \underline{0.0057} & \underline{0.0022} & \underline{0.0094} & \underline{0.0046}\\
\midrule
Males   & DMP-AP & 0.0160 & 0.0047 & 0.0324 & 0.0121\\
        & DMP-LC & \textbf{0.0078} & \textbf{0.0022} & \textbf{0.0100} & \textbf{0.0045}\\
        & CoDa      & 0.1784 & 0.1048 & 0.1624 & 0.0910\\
        & LC        & \underline{0.0087} & \underline{0.0034} & \underline{0.0129} & \underline{0.0069}\\
\bottomrule
\end{tabular}
\end{table}

Table~\ref{tab:fr_oos_by_cause_model_sexcols_1989} details the performance of OOS by cause. The picture is heterogeneous, but two patterns emerge. First, DMP-LC is consistently among the top performers in most causes for both sexes, while preserving the coherence between the total and cause decomposition by construction. In particular, DMP-LC is best for males in cardiovascular diseases (CVD), external causes (EXT), and residual category (OTHER) for RMSE and MAE, and it remains highly competitive for neoplasms (NEOP). For females, DMP-LC is best for EXT and OTHER, and near-best for CVD and infectious diseases (INF), confirming that a single latent time index can capture both the dominant trend and systematic reallocations across causes.

\renewcommand{\arraystretch}{0.86}
\begin{longtable}{@{}llrrrr@{}}
\caption{France -- Out-of-sample performance by cause and model, train from 1989, with sex shown as column groups. Lower is better for RMSE/MAE. Best in \textbf{bold}, second best \underline{underlined} (within each cause $\times$ sex, separately for RMSE and MAE).}\label{tab:fr_oos_by_cause_model_sexcols_1989}\\
\toprule
 &  & \multicolumn{2}{c}{\textbf{Females}} & \multicolumn{2}{c}{\textbf{Males}}\\
\cmidrule(lr){3-4}\cmidrule(lr){5-6}
\textbf{Cause} & \textbf{Model} & \textbf{RMSE} & \textbf{MAE} & \textbf{RMSE} & \textbf{MAE}\\
\midrule
\endfirsthead
\toprule
 &  & \multicolumn{2}{c}{\textbf{Females}} & \multicolumn{2}{c}{\textbf{Males}}\\
\cmidrule(lr){3-4}\cmidrule(lr){5-6}
\textbf{Cause} & \textbf{Model} & \textbf{RMSE} & \textbf{MAE} & \textbf{RMSE} & \textbf{MAE}\\
\midrule
\endhead
\midrule
\multicolumn{6}{r}{\small continued on next page}\\
\endfoot
\bottomrule
\endlastfoot

CVD & DMP-AP & 0.0752 & 0.0178 & 0.0117 & 0.0046\\
& DMP-LC & \underline{0.0055} & \underline{0.0023} & \textbf{0.0064} & \textbf{0.0029}\\
& CoDa      & \textbf{0.0038} & \textbf{0.0010} & 0.0129 & \underline{0.0039}\\
& LC        & 0.0094 & 0.0042 & \underline{0.0092} & 0.0045\\
\addlinespace

EXT & DMP-AP & 0.0217 & 0.0047 & 0.0046 & 0.0015\\
& DMP-LC & \textbf{0.0008} & \textbf{0.0003} & \textbf{0.0013} & \textbf{0.0004}\\
& CoDa      & 0.0017 & 0.0005 & 0.0016 & \underline{0.0005}\\
& LC        & \underline{0.0011} & \underline{0.0005} & \underline{0.0013} & 0.0005\\
\addlinespace

INF & DMP-AP & 0.0052 & 0.0012 & 0.0016 & 0.0006\\
& DMP-LC & \textbf{0.0004} & \underline{0.0002} & \underline{0.0013} & 0.0005\\
& CoDa      & 0.0007 & 0.0002 & 0.0017 & \underline{0.0005}\\
& LC        & \underline{0.0005} & \textbf{0.0002} & \textbf{0.0005} & \textbf{0.0002}\\
\addlinespace

NEOP & DMP-AP & 0.0462 & 0.0123 & 0.0085 & 0.0029\\
& DMP-LC & \underline{0.0008} & \underline{0.0003} & \textbf{0.0020} & \underline{0.0011}\\
& CoDa      & 0.0015 & 0.0007 & 0.0028 & \textbf{0.0010}\\
& LC        & \textbf{0.0007} & \textbf{0.0003} & \underline{0.0024} & 0.0015\\
\addlinespace

OTHER & DMP-AP & 0.0932 & 0.0210 & 0.0086 & 0.0029\\
& DMP-LC & \textbf{0.0033} & \textbf{0.0009} & \textbf{0.0022} & \textbf{0.0008}\\
& CoDa      & \underline{0.0038} & \underline{0.0012} & 0.0053 & 0.0016\\
& LC        & 0.0047 & 0.0014 & \underline{0.0031} & \underline{0.0011}\\
\addlinespace

RESP & DMP-AP & 0.0443 & 0.0080 & \textbf{0.0046} & \textbf{0.0014}\\
& DMP-LC & \textbf{0.0043} & \textbf{0.0011} & 0.0071 & \underline{0.0019}\\
& CoDa      & 0.0058 & 0.0015 & 0.0120 & 0.0033\\
& LC        & \underline{0.0043} & \underline{0.0012} & \underline{0.0069} & 0.0021\\
\end{longtable}

Second, the benchmark methods can be strong for specific causes but do not dominate uniformly. For example, CoDa is best for females in CVD (both RMSE and MAE) and performs competitively in several other female causes, while the classical LC benchmark is best for females in NEOP and for INF (notably on MAE), and it is also competitive for males in INF. As in the analysis of the U.S. data, such cause-by-cause gains should be interpreted alongside system-level coherence: independently fitted LC models (one per cause) do not guarantee that cause-specific forecasts sum to the forecast of total mortality, whereas both Bayesian specifications enforce coherence at each age and period.

The averages over causes in Table~\ref{tab:fr_oos_mean_across_causes_by_sex_1989} reinforce these conclusions. For males, DMP-LC attains the lowest mean RMSE and mean MAE across causes. For females, DMP-LC has the lowest mean RMSE, and CoDa attains the lowest mean MAE, while DMP-AP is an outlier due to very large errors in multiple causes and, especially, regarding the total.
\begin{table}[!htb]
\centering
\caption{France -- OOS mean performance across causes, train from 1989, by sex. Lower is better. Best in \textbf{bold}, second best \underline{underlined} (within each sex, separately for RMSE mean and MAE mean).}
\label{tab:fr_oos_mean_across_causes_by_sex_1989}
\small
\setlength{\tabcolsep}{50pt}
\renewcommand{\arraystretch}{1.15}
\begin{tabular}{@{}llrr@{}}
\toprule
\textbf{Sex} & \textbf{Model} & \textbf{RMSE mean} & \textbf{MAE mean}\\
\midrule
Female & DMP-AP & 0.0476 & 0.0108\\
        & DMP-LC & \textbf{0.0025} & \underline{0.0009}\\
        & CoDa      & \underline{0.0029} & \textbf{0.0008}\\
        & LC        & 0.0034 & 0.0013\\
\midrule
Male   & DMP-AP & 0.0066 & 0.0023\\
        & DMP-LC & \textbf{0.0034} & \textbf{0.0013}\\
        & CoDa      & 0.0061 & 0.0018\\
        & LC        & \underline{0.0039} & \underline{0.0016}\\
\bottomrule
\end{tabular}
\end{table}

\begin{figure}[!htb]
\centering
\includegraphics[width=1\linewidth]{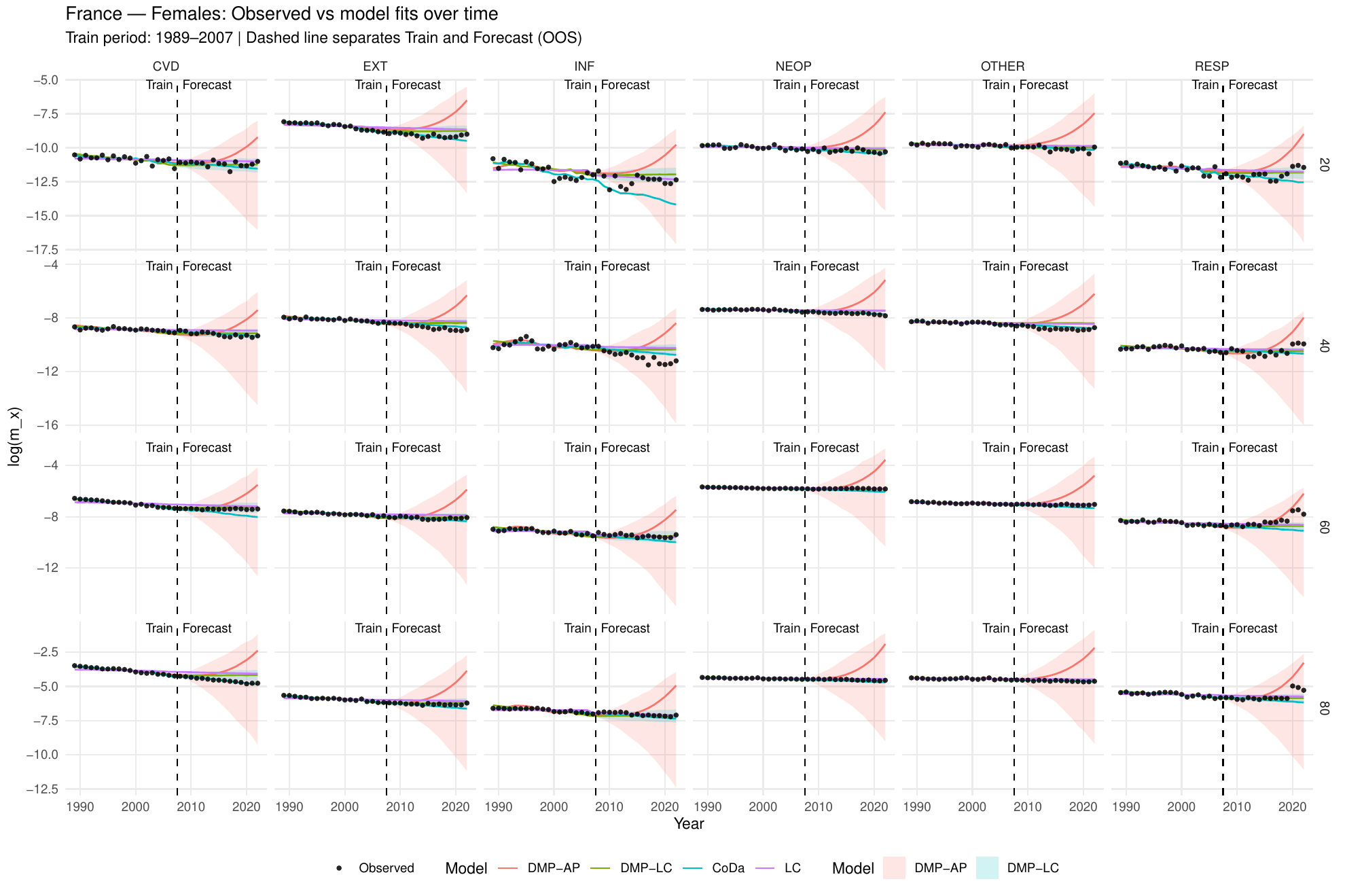}
\caption{France Female: Observed (black points) and fitted/forecast $\log$ mortality rates by cause for selected ages (20, 40, 60, 80). The vertical dashed line marks the end of the in-sample period (1989--2007) and the start of the out-of-sample forecasts.}
\label{fig:FRAFemale89}
\end{figure}

\begin{figure}[!htb]
\centering
\includegraphics[width=1\linewidth]{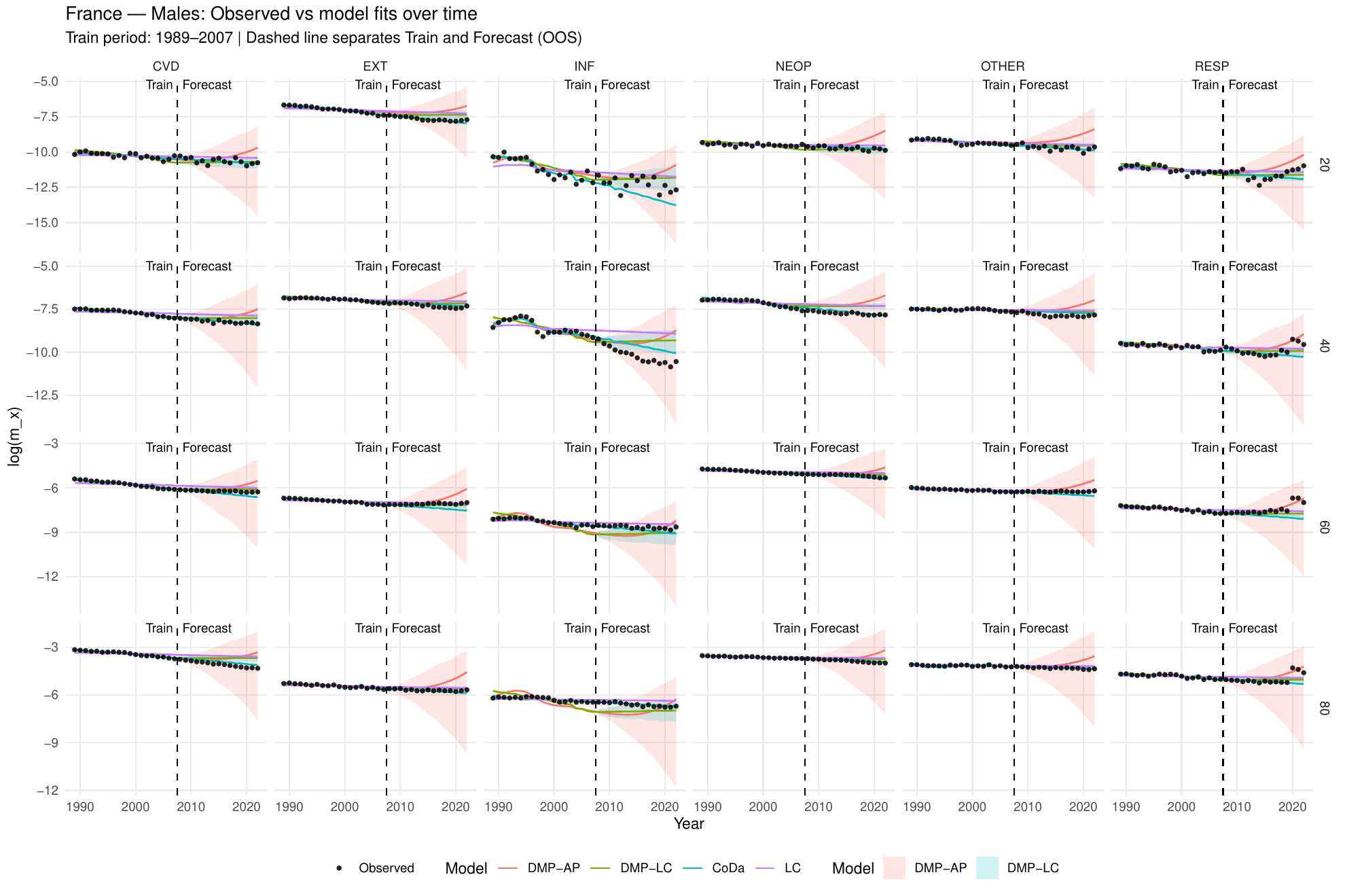}
\caption{France Male: Observed (black points) and fitted/forecast $\log$ mortality rates by cause for selected ages (20, 40, 60, 80). The vertical dashed line marks the end of the in-sample period (1989--2007) and the start of the out-of-sample forecasts.}
\label{fig:FRAMale89}
\end{figure}

Figures~\ref{fig:FRAFemale89} and~\ref{fig:FRAMale89} provide a visual check of these results by showing observed $\log$-mortality rates, fitted trajectories, and forecast behaviour for selected ages (20, 40, 60, 80), with the dashed line marking the transition from in-sample (1989--2007) to OOS forecasting (2008 onwards). The plots show that DMP-LC closely tracks the pre-2008 trends and produces stable extrapolations after 2007, with uncertainty bands that widen gradually with the forecast horizon. In contrast, DMP-AP displays very wide predictive intervals and noticeable drift in several panels (particularly for females), consistent with poor OOS performance on totals and across multiple causes. This behaviour is in line with a highly flexible period component whose second-order random-walk extrapolation can generate rapidly increasing uncertainty and curvature over longer horizons. CoDa and LC remain competitive in selected cause--age panels (notably INF and NEOP), but their strengths are not uniform across the full cause partition, and they do not provide the same combination of accurate totals, coherent decomposition, and joint probabilistic uncertainty for both level and composition delivered by the coherent Bayesian formulations.

In the Appendix \ref{Appendix:A}, Tables~\ref{tab:us_total_train_oos_by_sex_1979}--\ref{tab:fr_oos_mean_across_causes_by_sex_1979} show the evaluation using a longer in-sample window starting in 1979. For both countries, DMP-LC still provides the best performance on total mortality in both sexes, confirming the robustness of the LC-based coherent Bayesian specification when the estimation sample is expanded. In contrast, CoDa remains clearly less competitive overall in this rate-based evaluation, despite occasionally strong performance on specific causes.

For cause-specific performance, the longer in-sample window amplifies the heterogeneity already observed in the main text. In the U.S. data, CoDa often achieves the lowest averaged errors over causes (Table~\ref{tab:us_oos_mean_across_causes_by_sex_1979}), while DMP-AP/DMP-LC remain consistently competitive and typically dominate for totals. In France, the same pattern appears. Indeed, CoDa yields the lowest average errors across causes for females, whereas DMP-LC is the best overall for males and remains among the top performers by cause (Table~\ref{tab:fr_oos_mean_across_causes_by_sex_1979}). Overall, the appendix results reinforce the main conclusion: coherent Bayesian models, especially DMP-LC, offer the most reliable accuracy for total mortality while maintaining a coherent composition structure, whereas benchmark methods may excel for selected causes but do not match this system-level performance uniformly.

Overall, the estimation starting from 1979 confirms two robust findings: \begin{inparaenum}[1)]
\item DMP-LC provides the most stable and accurate total mortality forecasts across countries and sexes, and its advantage persists when the estimation window is substantially extended. 
\item Cause-performance remains heterogeneous and depends on the evaluation target. Specifically, CoDa can excel for several individual causes and can even minimise errors across causes in some contexts, such as the female population; nevertheless, it performs poorly on totals. 
\end{inparaenum}
This reinforces the proposed methodological approaches, emphasising that the coherent Bayesian joint models (DMP-AP and especially DMP-LC) deliver a stronger forecast--accurate totals, coherent decomposition, and a single probabilistic description of uncertainty. In contrast, benchmark approaches may provide cause-specific gains that do not generalise to the full competing-risks/compositional system.

\section{Conclusion}\label{sec:conclusion}

This paper introduces a Dirichlet–Multinomial–Poisson modelling framework for the analysis and forecasting of mortality by cause of death. The proposed hierarchical specification provides a modular structure in which the variation in total mortality by age and calendar year is modelled jointly with the allocation of deaths by causes. Randomness in both the total number of deaths and the cause-specific probabilities allows the model to capture age- and time-dependent dynamics in overall mortality and in the relative importance of different causes.

By modelling these components simultaneously, the framework delivers coherent forecasts of cause-specific mortality rates that are fully consistent with projections of all-cause mortality. Importantly, coherence is preserved not only at the level of point forecasts but also in the associated uncertainty, as dependence across causes is explicitly represented. The resulting approach offers a tractable and interpretable alternative to independent cause-specific models, while retaining analytical transparency and compatibility with Bayesian inference.

These benefits come at a cost. Full Bayesian estimation across the dimensions of age, period and cause can be computationally demanding and requires careful prior specification and diagnostic assessment, moreover, long-horizon projections may be sensitive to the assumed time structure. 
Empirically, the backtesting analysis for the United States and France highlights the importance of coherent joint modelling. The LC-based coherent Bayesian specification (DMP–LC) achieves the lowest or near-lowest out-of-sample error on the aggregate mortality surface across countries and sexes, while remaining stable over longer horizons. In particular, for France, the latent period index plays a pivotal role in capturing the dominant secular trend and borrowing strength across ages within a coherent multi-cause system. At the cause-specific level, DMP–LC attains low RMSE across most causes and populations, avoids systematic bias, and preserves the aggregation constraint linking causes to totals. Competing approaches may outperform in isolated cases, but they do not simultaneously guarantee coherence between aggregate and disaggregate forecasts, nor provide joint predictive distributions for totals and causes. When internal consistency and interpretable predictive uncertainty are central, the coherent Bayesian framework offers a favourable trade-off between accuracy, coherence, and uncertainty quantification.

Several extensions merit further investigation. First, the Dirichlet concentration parameter could be allowed to vary across age or period, enabling greater flexibility in the dispersion of the compositional component. Second, additional over-dispersion or cohort structure could be introduced to assess whether such features provide incremental predictive value beyond period dynamics. While both extensions are feasible, their adoption should be guided by a careful cost–benefit assessment: additional computational and modelling complexity should be justified by tangible improvements in calibration, coherence, or forecast accuracy in the intended applications.

\section*{Acknowledgment}

Andrea Nigri thanks the insightful comments and suggestions from seminar participants at the School of Demography, Australian National University, and at the Department of Actuarial Studies and Business Analytics, Macquarie University.

\newpage
\bibliographystyle{abbrvnat}
\bibliography{DMP_CoD}

\newpage

\appendix
\section{U.S. and France, 1979}\label{Appendix:A}

Tables~\ref{tab:us_total_train_oos_by_sex_1979}--\ref{tab:us_oos_mean_across_causes_by_sex_1979} and
\ref{tab:fr_total_train_oos_by_sex_1979}--\ref{tab:fr_oos_mean_across_causes_by_sex_1979} report a robustness check in which the in-sample window is extended back to 1979 (keeping the same $15$-year OOS horizon). This experiment is informative because the longer estimation sample spans additional epidemiological phases and structural shifts, thereby stress-testing the extrapolation behaviour of the competing models.
\begin{table}[!htb]
\centering
\caption{U.S. -- Total mortality performance (Train vs OOS) by sex, train from 1979. Lower is better. Best in \textbf{bold}, second best \underline{underlined} (within each sex, separately for each metric).}
\label{tab:us_total_train_oos_by_sex_1979}
\small
\setlength{\tabcolsep}{29pt}
\renewcommand{\arraystretch}{1.15}
\begin{tabular}{@{}llrrrr@{}}
\toprule
            &           & \multicolumn{2}{c}{Train} & \multicolumn{2}{c}{OOS} \\
\cmidrule(lr){3-4}\cmidrule(lr){5-6}
\textbf{Sex} & \textbf{Model} & \textbf{RMSE} & \textbf{MAE} & \textbf{RMSE} & \textbf{MAE}\\
\midrule
Females & DMP-AP & 0.0132 & 0.0039 & \underline{0.0184} & \underline{0.0069}\\
        & DMP-LC & \textbf{0.0044} & \textbf{0.0014} & \textbf{0.0101} & \textbf{0.0042}\\
        & CoDa      & 0.1446 & 0.0831 & 0.1458 & 0.0804\\
        & LC        & \underline{0.0070} & \underline{0.0030} & 0.0406 & 0.0178\\
\midrule
Males   & DMP-AP & \underline{0.0254} & \underline{0.0084} & \underline{0.0290} & \underline{0.0102}\\
        & DMP-LC & \textbf{0.0046} & \textbf{0.0017} & \textbf{0.0127} & \textbf{0.0058}\\
        & CoDa      & 0.1657 & 0.1016 & 0.1668 & 0.0951\\
        & LC        & 0.1177 & 0.0257 & 0.0565 & 0.0219\\
\bottomrule
\end{tabular}
\end{table}

\begin{table}[!htb]
\centering
\caption{France -- Total mortality performance (Train vs OOS) by sex, train from 1979. Lower is better. Best in \textbf{bold}, second best \underline{underlined} (within each sex, separately for each metric).}
\label{tab:fr_total_train_oos_by_sex_1979}
\small
\setlength{\tabcolsep}{29pt}
\renewcommand{\arraystretch}{1.15}
\begin{tabular}{@{}llrrrr@{}}
\toprule
    &   & \multicolumn{2}{c}{Train} & \multicolumn{2}{c}{OOS} \\
\cmidrule(lr){3-4}\cmidrule(lr){5-6} 
\textbf{Sex} & \textbf{Model} & \textbf{RMSE} & \textbf{MAE} & \textbf{RMSE} & \textbf{MAE}\\
\midrule
Females & DMP-AP & 0.0175 & 0.0048 & 0.1376 & 0.0354\\
        & DMP-LC & \textbf{0.0040} & \textbf{0.0012} & \textbf{0.0074} & \textbf{0.0028}\\
        & CoDa      & 0.1594 & 0.0894 & 0.1403 & 0.0730\\
        & LC        & \underline{0.0127} & \underline{0.0056} & \underline{0.1011} & \underline{0.0259}\\
\midrule
Males   & DMP-AP & 0.0215 & 0.0068 & 0.0247 & 0.0100\\
        & DMP-LC & \textbf{0.0082} & \textbf{0.0023} & \textbf{0.0108} & \textbf{0.0047}\\
        & CoDa      & 0.1869 & 0.1114 & 0.1603 & 0.0889\\
        & LC        & \underline{0.0161} & \underline{0.0072} & \underline{0.0270} & \underline{0.0157}\\
\bottomrule
\end{tabular}
\end{table}

Across both countries and sexes, the rankings by total are remarkably stable. Indeed, DMP-LC remains overall the best-performing method in-sample and out-of-sample. For the U.S. (Table~\ref{tab:us_total_train_oos_by_sex_1979}), DMP-LC attains the lowest OOS errors for both females and males, with DMP-AP generally second-best and the benchmarks (LC and especially CoDa) substantially less competitive on total rates. France shows the same ordering for males (Table~\ref{tab:fr_total_train_oos_by_sex_1979}). For French females, DMP-LC again dominates, while DMP-AP and LC exhibit much larger OOS errors, indicating that the additive AP specification can remain sensitive in this setting even when trained on a longer historical window. Overall, extending the in-sample period strengthens the interpretation that the LC-based coherent Bayesian specification provides the most reliable system-level forecasts, in the sense of simultaneously modelling the level and maintaining a coherent decomposition.

At the cause level, the longer in-sample window accentuates the heterogeneity already visible in the main text. In the U.S. (Table~\ref{tab:us_oos_by_cause_model_sexcols_1979}), CoDa and LC can be best for specific causes (e.g., INF often favours CoDa/LC; EXT can favour LC for females), while DMP-LC is frequently competitive and often best for key male causes such as CVD. DMP-AP remains close to the best methods in several cells (especially for CVD females), suggesting that an additive age--period structure can still capture important movements in the composition, but its system-level advantage is not as robust as DMP-LC when the target is accurate totals plus coherence.


\begin{longtable}{@{}llrrrr@{}}
\caption{U.S. -- Out-of-sample performance by cause and model, train from 1979, with sex shown as column groups. Lower is better for RMSE/MAE. Best in \textbf{bold}, second best \underline{underlined} (within each cause $\times$ sex, separately for RMSE and MAE).}\label{tab:us_oos_by_cause_model_sexcols_1979}\\

\toprule
 &  & \multicolumn{2}{c}{\textbf{Females}} & \multicolumn{2}{c}{\textbf{Males}}\\
\cmidrule(lr){3-4}\cmidrule(lr){5-6}
\textbf{Cause} & \textbf{Model} & \textbf{RMSE} & \textbf{MAE} & \textbf{RMSE} & \textbf{MAE}\\
\midrule
\endfirsthead

\toprule
 &  & \multicolumn{2}{c}{\textbf{Females}} & \multicolumn{2}{c}{\textbf{Males}}\\
\cmidrule(lr){3-4}\cmidrule(lr){5-6}
\textbf{Cause} & \textbf{Model} & \textbf{RMSE} & \textbf{MAE} & \textbf{RMSE} & \textbf{MAE}\\
\midrule
\endhead

\midrule
\multicolumn{6}{r}{\small continued on next page}\\
\endfoot

\bottomrule
\endlastfoot

CVD & DMP-AP & \textbf{0.0049} & 0.0021 & 0.0177 & 0.0057\\
& DMP-LC & 0.0142 & 0.0059 & \textbf{0.0109} & 0.0050\\
& CoDa      & \underline{0.0063} & \textbf{0.0019} & \underline{0.0145} & \textbf{0.0041}\\
& LC        & 0.0263 & 0.0126 & 0.0233 & 0.0127\\
\addlinespace

EXT & DMP-AP & 0.0005 & 0.0002 & \underline{0.0007} & \underline{0.0004}\\
& DMP-LC & \underline{0.0004} & \underline{0.0002} & \textbf{0.0005} & \textbf{0.0003}\\
& CoDa      & 0.0007 & 0.0003 & 0.0008 & 0.0004\\
& LC        & \textbf{0.0003} & \textbf{0.0002} & 0.0007 & 0.0004\\
\addlinespace

INF & DMP-AP & 0.0010 & 0.0003 & \underline{0.0010} & \underline{0.0005}\\
& DMP-LC & 0.0018 & 0.0007 & 0.0018 & 0.0008\\
& CoDa      & \textbf{0.0002} & \textbf{0.0001} & \underline{0.0007} & \underline{0.0002}\\
& LC        & \underline{0.0004} & \underline{0.0001} & \textbf{0.0004} & \textbf{0.0002}\\
\addlinespace

NEOP & DMP-AP & 0.0026 & 0.0012 & \underline{0.0027} & \underline{0.0013}\\
& DMP-LC & 0.0012 & 0.0006 & 0.0028 & 0.0015\\
& CoDa      & \textbf{0.0006} & \textbf{0.0003} & \textbf{0.0022} & \textbf{0.0010}\\
& LC        & \underline{0.0009} & \underline{0.0006} & 0.0030 & 0.0019\\
\addlinespace

OTHER & DMP-AP & \underline{0.0094} & 0.0037 & 0.0109 & 0.0038\\
& DMP-LC & 0.0131 & 0.0047 & \underline{0.0091} & \underline{0.0037}\\
& CoDa      & \textbf{0.0060} & \textbf{0.0021} & \textbf{0.0086} & \textbf{0.0031}\\
& LC        & 0.0126 & \underline{0.0035} & 0.0613 & 0.0251\\
\addlinespace

RESP & DMP-AP & 0.0073 & 0.0024 & \underline{0.0048} & \underline{0.0018}\\
& DMP-LC & 0.0059 & 0.0021 & 0.0062 & 0.0025\\
& CoDa      & \textbf{0.0031} & \textbf{0.0010} & \textbf{0.0043} & \textbf{0.0016}\\
& LC        & \underline{0.0048} & \underline{0.0019} & 0.0058 & 0.0018\\
\end{longtable}

France exhibits a similar pattern in Table~\ref{tab:fr_oos_by_cause_model_sexcols_1979}. For males, DMP-LC is again broadly strong across causes and is best in several major cells (e.g., CVD and OTHER). For females, CoDa can be extremely competitive in some causes (notably CVD and OTHER), as reflected in the averages across causes. However, this cause-level strength does not translate into good total-mortality performance in a rate-based evaluation, reinforcing a key point: methods can optimise different aspects of the problem (level versus composition), and superior cause-level fit does not automatically imply superior system-level coherence/accuracy on totals.

\renewcommand{\arraystretch}{0.96}
\begin{longtable}{@{}llrrrr@{}}
\caption{France -- Out-of-sample performance by cause and model, train from 1979, with sex shown as column groups. Lower is better for RMSE/MAE. Best in \textbf{bold}, second best \underline{underlined} (within each cause $\times$ sex, separately for RMSE and MAE).}
\label{tab:fr_oos_by_cause_model_sexcols_1979}\\
\toprule
 &  & \multicolumn{2}{c}{\textbf{Females}} & \multicolumn{2}{c}{\textbf{Males}}\\
\cmidrule(lr){3-4}\cmidrule(lr){5-6}
\textbf{Cause} & \textbf{Model} & \textbf{RMSE} & \textbf{MAE} & \textbf{RMSE} & \textbf{MAE}\\
\midrule
\endfirsthead
\toprule
 &  & \multicolumn{2}{c}{\textbf{Females}} & \multicolumn{2}{c}{\textbf{Males}}\\
\cmidrule(lr){3-4}\cmidrule(lr){5-6}
\textbf{Cause} & \textbf{Model} & \textbf{RMSE} & \textbf{MAE} & \textbf{RMSE} & \textbf{MAE}\\
\midrule
\endhead
\midrule
\multicolumn{6}{r}{\small continued on next page}\\
\endfoot
\bottomrule
\endlastfoot

CVD & DMP-AP & 0.0466 & 0.0123 & 0.0122 & 0.0045\\
& DMP-LC & \underline{0.0058} & \underline{0.0025} & \textbf{0.0069} & \textbf{0.0030}\\
& CoDa      & \textbf{0.0031} & \textbf{0.0010} & 0.0154 & \underline{0.0049}\\
& LC        & 0.0229 & 0.0110 & \underline{0.0227} & 0.0119\\
\addlinespace

EXT & DMP-AP & 0.0105 & 0.0024 & 0.0022 & 0.0008\\
& DMP-LC & \underline{0.0015} & \underline{0.0006} & \textbf{0.0014} & \textbf{0.0005}\\
& CoDa      & \textbf{0.0013} & \textbf{0.0005} & \underline{0.0021} & \underline{0.0007}\\
& LC        & 0.0039 & 0.0018 & 0.0021 & 0.0013\\
\addlinespace

INF & DMP-AP & 0.0027 & 0.0007 & 0.0014 & 0.0005\\
& DMP-LC & \textbf{0.0006} & \textbf{0.0002} & \underline{0.0007} & \underline{0.0003}\\
& CoDa      & \underline{0.0015} & \underline{0.0005} & 0.0020 & 0.0006\\
& LC        & 0.0982 & 0.0139 & \textbf{0.0006} & \textbf{0.0003}\\
\addlinespace

NEOP & DMP-AP & 0.0209 & 0.0064 & 0.0050 & 0.0020\\
& DMP-LC & 0.0013 & \underline{0.0006} & 0.0028 & \underline{0.0014}\\
& CoDa      & \underline{0.0010} & \textbf{0.0006} & \textbf{0.0027} & \textbf{0.0009}\\
& LC        & \textbf{0.0010} & 0.0006 & \underline{0.0038} & 0.0024\\
\addlinespace

OTHER & DMP-AP & 0.0459 & 0.0114 & 0.0064 & 0.0025\\
& DMP-LC & 0.0058 & 0.0019 & \textbf{0.0029} & \textbf{0.0010}\\
& CoDa      & \textbf{0.0026} & \textbf{0.0009} & 0.0079 & 0.0029\\
& LC        & \underline{0.0092} & \underline{0.0033} & \underline{0.0042} & \underline{0.0020}\\
\addlinespace

RESP & DMP-AP & 0.0128 & 0.0026 & \textbf{0.0059} & \textbf{0.0015}\\
& DMP-LC & \textbf{0.0042} & \textbf{0.0012} & 0.0073 & \underline{0.0020}\\
& CoDa      & 0.0065 & \underline{0.0017} & 0.0097 & 0.0026\\
& LC        & \underline{0.0044} & 0.0017 & \underline{0.0072} & 0.0026\\
\end{longtable}

Tables~\ref{tab:us_oos_mean_across_causes_by_sex_1979} and~\ref{tab:fr_oos_mean_across_causes_by_sex_1979} further clarify these trade-offs. In the U.S., CoDa attains the lowest mean RMSE/MAE for females and is also best on the mean metrics for males, while DMP-AP/DMP-LC remain close competitors. In France, CoDa yields the lowest mean errors across causes for females, whereas DMP-LC is best overall for males. These results suggest that when one aggregates uniformly across causes, compositional methods can excel; nevertheless, the totals tell a different story, and the coherent Bayesian LC formulation is the only approach that is consistently strong on the full system (accurate totals plus a coherent decomposition).

\begin{table}[!htb]
\centering
\caption{U.S. -- OOS mean performance across causes, train from 1979, by sex. Lower is better. Best in \textbf{bold}, second best \underline{underlined} (within each sex, separately for RMSE mean and MAE mean).}
\label{tab:us_oos_mean_across_causes_by_sex_1979}
\small
\setlength{\tabcolsep}{50pt}
\renewcommand{\arraystretch}{1.15}
\begin{tabular}{@{}llrr@{}}
\toprule
\textbf{Sex} & \textbf{Model} & \textbf{RMSE mean} & \textbf{MAE mean}\\
\midrule
Females & DMP-AP & \underline{0.0043} & \underline{0.0017}\\
        & DMP-LC & 0.0061 & 0.0024\\
        & CoDa      & \textbf{0.0028} & \textbf{0.0010}\\
        & LC        & 0.0076 & 0.0031\\
\midrule
Males   & DMP-AP & 0.0063 & \underline{0.0023}\\
        & DMP-LC & \underline{0.0052} & 0.0023\\
        & CoDa      & \textbf{0.0052} & \textbf{0.0017}\\
        & LC        & 0.0158 & 0.0070\\
\bottomrule
\end{tabular}
\end{table}

\begin{table}[!htb]
\centering
\caption{France -- OOS mean performance across causes, train from 1979, by sex. Lower is better. Best in \textbf{bold}, second best \underline{underlined} (within each sex, separately for RMSE mean and MAE mean).}
\label{tab:fr_oos_mean_across_causes_by_sex_1979}
\small
\setlength{\tabcolsep}{50pt}
\renewcommand{\arraystretch}{1.15}
\begin{tabular}{@{}llrr@{}}
\toprule
\textbf{Sex} & \textbf{Model} & \textbf{RMSE mean} & \textbf{MAE mean}\\
\midrule
Females & DMP-AP & \underline{0.0232} & 0.0060\\
        & DMP-LC & 0.0032 & \underline{0.0012}\\
        & CoDa      & \textbf{0.0027} & \textbf{0.0009}\\
        & LC        & 0.0233 & 0.0054\\
\midrule
Males   & DMP-AP & \underline{0.0055} & 0.0020\\
        & DMP-LC & \textbf{0.0037} & \textbf{0.0014}\\
        & CoDa      & 0.0066 & \underline{0.0021}\\
        & LC        & 0.0067 & 0.0034\\
\bottomrule
\end{tabular}
\end{table}

\begin{figure}[!htb]
\centering
\includegraphics[width=0.83\linewidth]{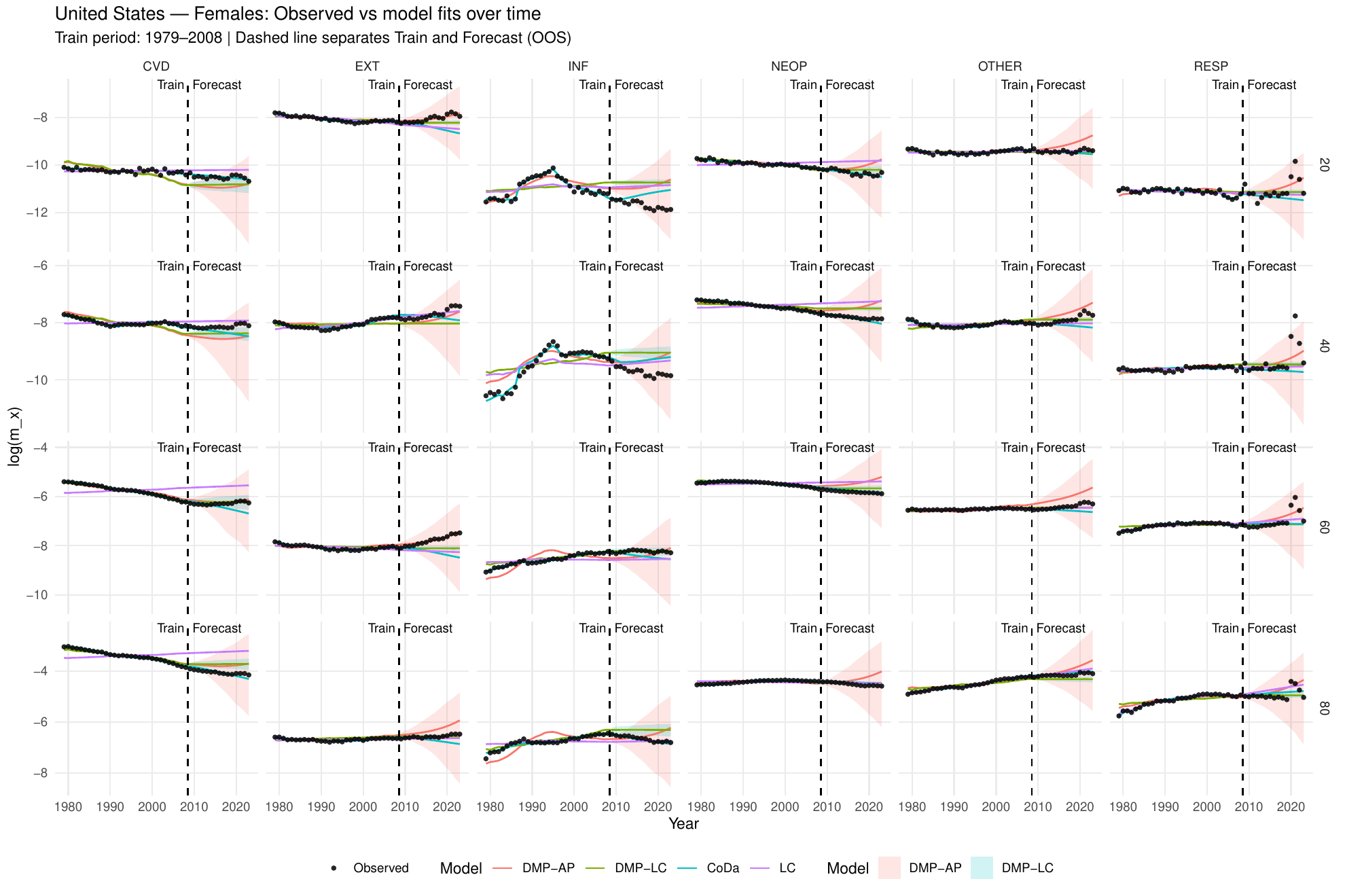}
\caption{U.S. Female: Observed (black points) and fitted/forecast $\log$ mortality rates by cause for selected ages (20, 40, 60, 80). The vertical dashed line marks the end of the in-sample period (1979--2008) and the start of the out-of-sample forecasts.}
\label{fig:USAFemale79}
\end{figure}

\begin{figure}[!htb]
\centering
\includegraphics[width=0.83\linewidth]{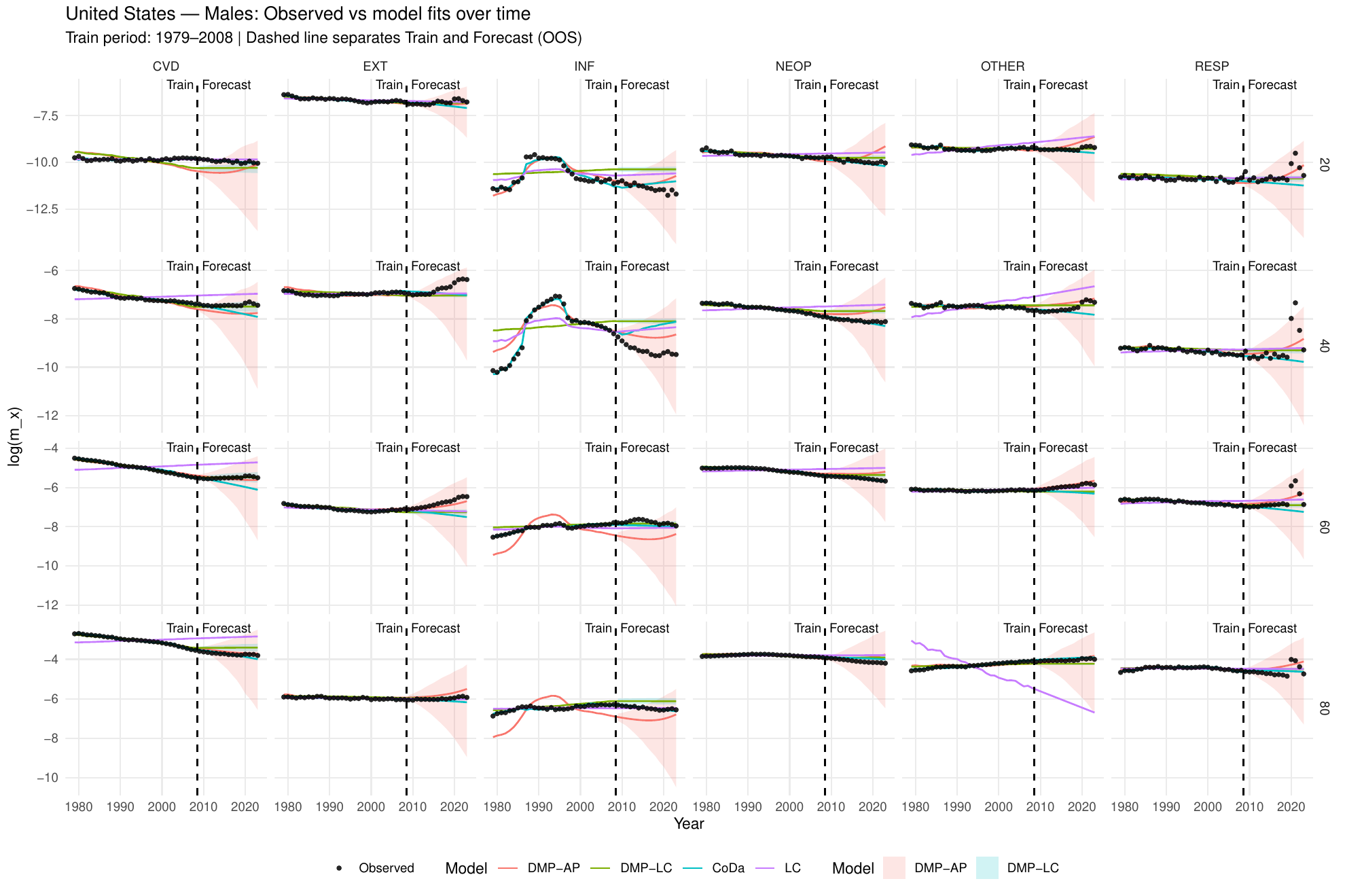}
\caption{U.S. Male: Observed (black points) and fitted/forecast $\log$ mortality rates by cause for selected ages (20, 40, 60, 80). The vertical dashed line marks the end of the in-sample period (1979--2008) and the start of the out-of-sample forecasts.}
\label{fig:USAMale79}
\end{figure}

\begin{figure}[!htb]
\centering
\includegraphics[width=0.83\linewidth]{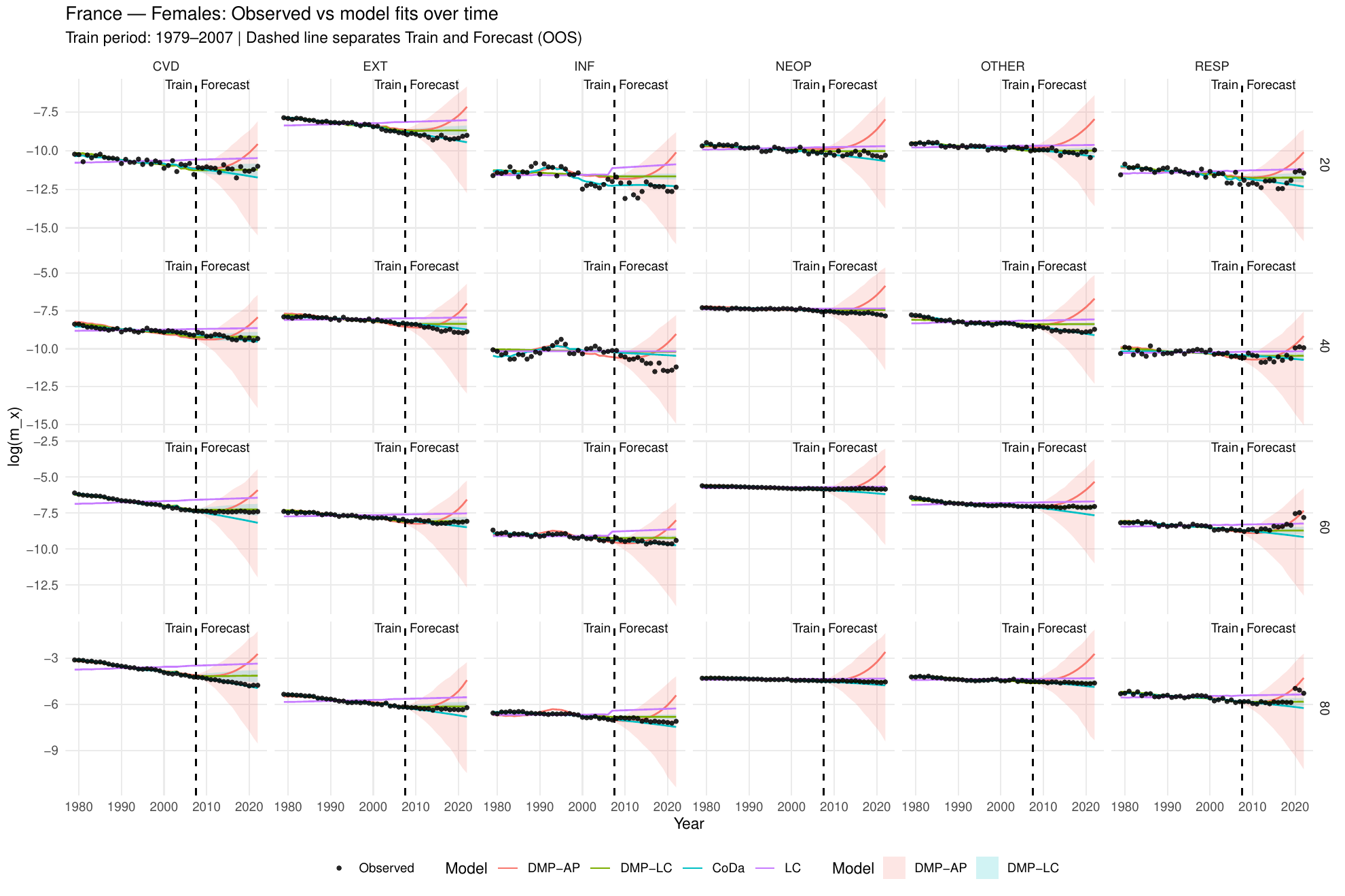}
\caption{France Female: Observed (black points) and fitted/forecast $\log$ mortality rates by cause for selected ages (20, 40, 60, 80). The vertical dashed line marks the end of the in-sample period (1979--2007) and the start of the out-of-sample forecasts.}
\label{fig:FRAFemale79}
\end{figure}

\begin{figure}[!htb]
\centering
\includegraphics[width=0.83\linewidth]{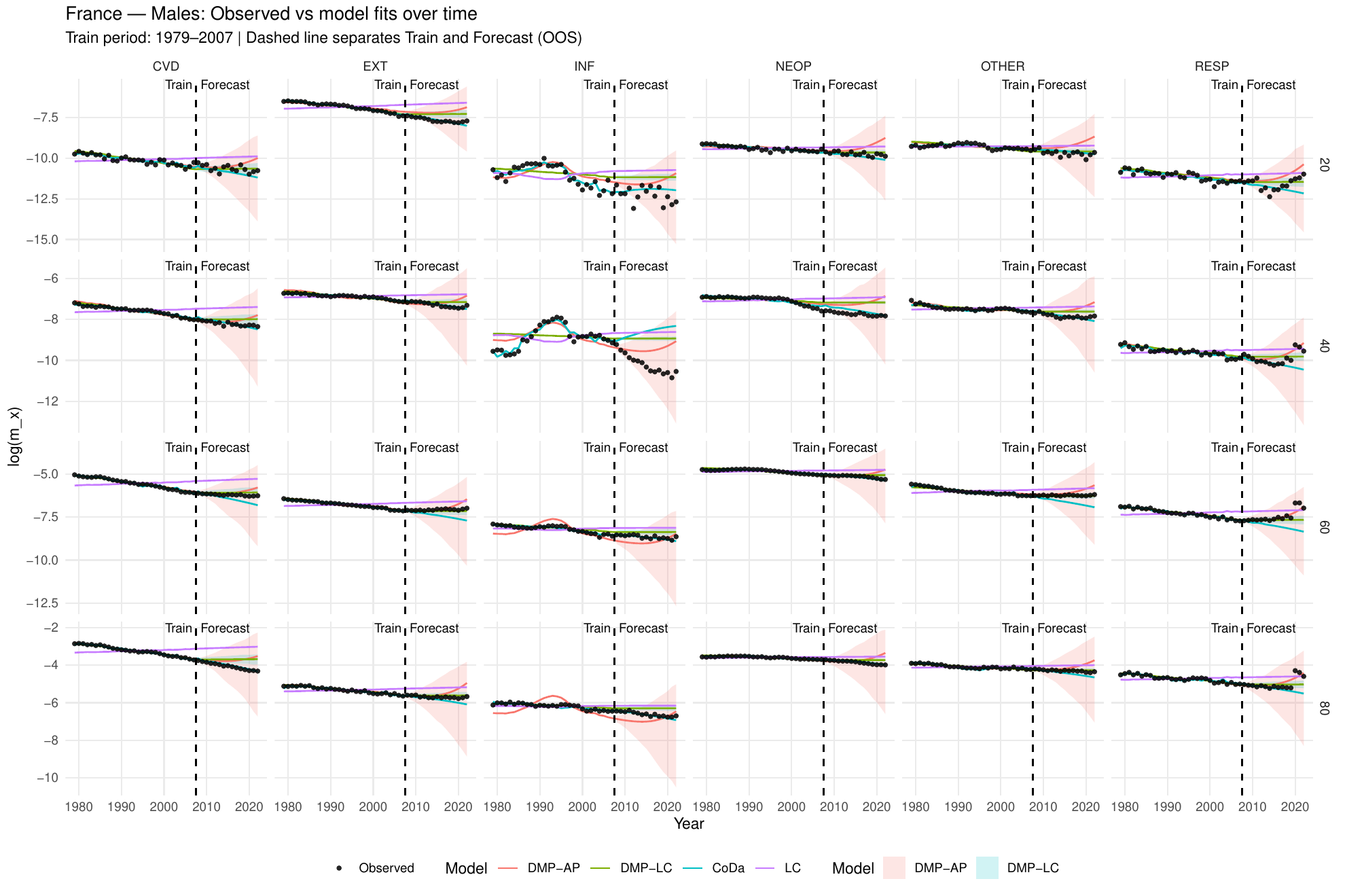}
\caption{France Male: Observed (black points) and fitted/forecast $\log$ mortality rates by cause for selected ages (20, 40, 60, 80). The vertical dashed line marks the end of the in-sample period (1979--2007) and the start of the out-of-sample forecasts.}\label{fig:FRAMale79}
\end{figure}

Figures~\ref{fig:USAFemale79}-\ref{fig:FRAMale79} provide visual checks of these results by showing observed $\log$ mortality rates, fitted trajectories, and forecast behaviour for selected ages (20, 40, 60, 80), with the dashed line marking the transition from in-sample (1979-2008) to OOS forecasting (2008 onward).

\newpage

\section{Benchmark models: LC and CoDa}\label{app:benchmarks_lc_coda}

We summarise the two benchmark models used for comparison, i.e. the LC model fitted to cause-specific mortality rates, and a CoDa approach fitted to life-table death distributions. Our implementation follows the specifications in \citet{kjaergaard2019forecasting}. 

\subsection{Cause-by-cause LC}\label{app:lc}

Let $D_{t,x,c}$ denote observed deaths in year $t$, age group $x$, cause $c$, and let $m_{t,x,c}$ be the corresponding (cause-specific) death rate. The LC benchmark is fitted \emph{separately for each cause} using the standard bilinear structure
\begin{equation*}
\log m_{t,x,c} \;=\; \alpha_{x,c} \;+\; \beta_{x,c}\,\kappa_{t,c} \;+\; \varepsilon_{t,x,c},
\end{equation*}
where $\alpha_{x,c}$ is the average age profile for cause $c$, $\kappa_{t,c}$ is the time index, and $\beta_{x,c}$ measures the age-specific sensitivity to changes in $\kappa_{t,c}$ (with standard identifiability constraints such as $\sum_x \beta_{x,c}=1$ and $\sum_t \kappa_{t,c}=0$). Forecasts are obtained by extrapolating $\kappa_{t,c}$ (typically with autoregressive integrated moving average (ARIMA)-type time-series models) and then reconstructing $\widehat{m}_{t,x,c}$.

Because LC is fitted independently by cause, coherence is not guaranteed: in general, $\sum_c \widehat{m}_{t,x,c}$ need not match the independently modelled total. Therefore, we treat LC as a strong but potentially non-coherent baseline \citep{wilmoth1995mortality}.

\subsection{CoDa on life-table death counts}\label{app:coda}

The CoDa benchmark models the distribution of deaths rather than the cause-specific rates. Starting from a multiple-decrement life table, define all-cause life-table deaths $d_{t,x}$ (for a given radix, such as $10^5$) and transform observed cause-of-death counts into cause-specific life-table deaths via
\begin{equation*}
d_{t,x,c} \;=\; d_{t,x}\,\frac{D_{t,x,c}}{D_{t,x}}, 
\qquad\text{where}\qquad
D_{t,x}=\sum_{c=1}^K D_{t,x,c}.
\end{equation*}
This produces, for each year $t$, a nonnegative array $\{d_{t,x,c}\}$ whose elements sum to the radix, i.e.\ a composition over the age--cause cells. 

To work in an unconstrained space, CoDa applies the centred $\log$-ratio (CLR) transformation. If we stack the age--cause cells in a single index $j \in \{1,\dots,NK\}$, where $N$ is the number of age groups and $K$ is the number of causes, and write $d_{t,j}$, the CLR is
\begin{equation*}
\operatorname{clr}(d_{t,j}) \;=\; \log\!\left(\frac{d_{t,j}}{g_t}\right),
\qquad
g_t \;=\; \Big(\prod_{j=1}^{NK} d_{t,j}\Big)^{1/(NK)}.
\end{equation*}
The transformed series is then decomposed by singular value decomposition, yielding an LC-type factorisation on the CLR scale:
\begin{equation*}
\operatorname{clr}(d_{t,j}) \;=\; \alpha_j \;+\; \sum_{p=1}^{P} \beta_{j,p}\,k_{t,p} \;+\; \varepsilon_{t,j}.
\end{equation*}
Forecasts are produced by time-series extrapolation (e.g.\ ARIMA) of the score vectors $k_{t,p}$, followed by an inverse-CLR map and closure back to the simplex (so that the predicted death distribution sums to the radix). In the notation of compositional operations, the back-transformation can be written schematically as
\begin{equation*}
\widehat{\mathbf d}_{t} \;=\; \operatorname{clr}^{-1}\!\Big(\sum_{p=1}^{P} \boldsymbol{\beta}_{p}\,\widehat{k}_{t,p}\Big)\;\oplus\;\boldsymbol{\alpha},
\end{equation*}
where $\boldsymbol{\alpha}$ is the baseline (time-average) age–cause death composition, and $\operatorname{clr}^{-1}(\cdot)$ includes the closure operator that ensures the sum constraint.

\newpage

\section{Posterior summary}\label{App:C}

In Tables~\ref{tab:fr_posteriors_scalars_all} and~\ref{tab:us_posteriors_scalars_all}, we present posterior summaries for scalar parameters of the Bayesian AP and LC models, by sex and in-sample years for France and the U.S., respectively.
\setlength{\tabcolsep}{3.7pt}
\renewcommand{\arraystretch}{1.08}
\footnotesize
\begin{longtable}{@{}llrrrrrrrrrrrrrr@{}}
\caption{France -- Posterior summary for scalar parameters (DMP-AP and DMP-LC), by sex and in-sample start year (1979 vs 1989).}
\label{tab:fr_posteriors_scalars_all}\\
\toprule
Year & \multicolumn{2}{l}{Parameter} & \multicolumn{5}{c}{Males} & \multicolumn{7}{c}{Females} \\
& & Mean & SD & 5\% & 50\% & 95\% & $n_{\mathrm{eff}}$ & $\widehat{R}$ & Mean & SD & 5\% & 50\% & 95\% & $n_{\mathrm{eff}}$ & $\widehat{R}$ \\
\midrule
\endfirsthead
\toprule
Year & \multicolumn{2}{l}{Parameter} & \multicolumn{5}{c}{Males} & \multicolumn{7}{c}{Females} \\
& & Mean & SD & 5\% & 50\% & 95\% & $n_{\mathrm{eff}}$ & $\widehat{R}$ & Mean & SD & 5\% & 50\% & 95\% & $n_{\mathrm{eff}}$ & $\widehat{R}$ \\
\midrule
\endhead
\midrule
\multicolumn{15}{r}{\small continued on next page}\\
\endfoot
\bottomrule
\endlastfoot
\multicolumn{14}{l}{\hspace{-.06in} \textbf{\underline{DMP-AP}}}\\
1979 & $\nu_0$           & -4.72  & 0.00  & -4.72 & -4.72 & -4.71 & 2923 & 1.00  & -5.36  & 0.00  & -5.36 & -5.36 & -5.36 & 4465 & 1.00 \\
& $\sigma_{\delta}$  & 0.67   & 0.11  & 0.52  & 0.65  & 0.86  & 667  & 1.01 & 0.60   & 0.10  & 0.46  & 0.59  & 0.78  & 178  & 1.00 \\
& $\sigma_{\pi}$     & 0.03   & 0.00  & 0.03  & 0.03  & 0.04  & 860  & 1.01 & 0.05   & 0.01  & 0.04  & 0.05  & 0.07  & 204  & 1.02 \\
& $\sigma_{\zeta}$       & 0.32   & 0.02  & 0.28  & 0.32  & 0.36  & 618  & 1.00 & 0.28   & 0.02  & 0.25  & 0.28  & 0.32  & 211  & 1.01 \\
& $\sigma_{\lambda}$       & 0.02   & 0.00  & 0.02  & 0.02  & 0.03  & 508  & 1.01 & 0.02   & 0.00  & 0.01  & 0.02  & 0.02  & 173  & 1.03 \\
& $\phi$             & 652.72 & 21.87 & 616.82& 652.82& 688.93& 684  & 1.00 & 898.07 & 34.90 & 844.81& 895.57& 958.71& 167  & 1.03 \\
\addlinespace

\multicolumn{14}{l}{\hspace{-.06in}\textbf{\underline{DMP-LC}}}\\
1979 & $\nu_0$           & -4.73  & 0.00  & -4.73 & -4.73 & -4.73 & 3798 & 1.00 & -5.38  & 0.00  & -5.38 & -5.38 & -5.38 & 3106 & 1.00 \\
& $\sigma_{\kappa}$  & 0.65   & 0.09  & 0.53  & 0.64  & 0.81  & 522  & 1.00 & 0.81   & 0.10  & 0.66  & 0.80  & 0.99  & 445  & 1.01 \\
& $\sigma_{\zeta}$       & 0.31   & 0.02  & 0.27  & 0.30  & 0.35  & 438  & 1.00 & 0.28   & 0.02  & 0.24  & 0.27  & 0.31  & 387  & 1.01 \\
& $\phi$             & 390.58 & 11.25 & 372.88& 390.25& 409.77& 597  & 1.01 & 667.74 & 23.44 & 629.35& 667.30& 707.14& 268  & 1.01 \\
\addlinespace

\multicolumn{8}{l}{\hspace{-.06in}\textbf{\underline{DMP-AP}}}\\
1989 & $\nu_0$           & -4.83   & 0.00  & -4.83  & -4.83  & -4.83  & 4608 & 1.00 & -5.48    & 0.00   & -5.48   & -5.48   & -5.48   & 5609 & 1.00 \\
& $\sigma_{\delta}$  & 0.66    & 0.11  & 0.51   & 0.64   & 0.87   & 631  & 1.00 & 0.60     & 0.11   & 0.46    & 0.58    & 0.80    & 366  & 1.02 \\
& $\sigma_{\pi}$     & 0.04    & 0.01  & 0.03   & 0.04   & 0.05   & 850  & 1.01 & 0.06     & 0.01   & 0.05    & 0.06    & 0.08    & 412  & 1.00 \\
& $\sigma_{\zeta}$       & 0.34    & 0.03  & 0.30   & 0.34   & 0.39   & 711  & 1.00 & 0.31     & 0.02   & 0.28    & 0.31    & 0.35    & 496  & 1.00 \\
& $\sigma_{\lambda}$       & 0.03    & 0.00  & 0.03   & 0.03   & 0.04   & 545  & 1.00 & 0.03     & 0.00   & 0.02    & 0.03    & 0.03    & 379  & 1.01 \\
& $\phi$             & 839.40  & 36.75 & 781.70 & 838.02 & 902.39 & 595  & 1.00 & 2032.95  & 130.66 & 1832.24 & 2027.17 & 2256.96 & 433  & 1.01 \\
\addlinespace

\multicolumn{8}{l}{\hspace{-.06in}\textbf{\underline{DMP-LC}}}\\
1989 & $\nu_0$           & -4.84   & 0.00  & -4.84  & -4.84  & -4.84  & 2400 & 1.00 & -5.49    & 0.00   & -5.49   & -5.49   & -5.49   & 4425 & 1.00 \\
& $\sigma_{\kappa}$  & 0.75    & 0.12  & 0.58   & 0.74   & 0.97   & 285  & 1.01 & 0.89     & 0.13   & 0.71    & 0.88    & 1.12    & 447  & 1.00 \\
& $\sigma_{\zeta}$       & 0.34    & 0.03  & 0.30   & 0.34   & 0.39   & 229  & 1.00 & 0.31     & 0.02   & 0.27    & 0.30    & 0.35    & 350  & 1.01 \\
& $\phi$             & 634.99  & 24.50 & 593.58 & 635.33 & 675.55 & 194  & 1.01 & 1316.73  & 74.93  & 1195.37 & 1316.18 & 1441.40 & 610  & 1.00 \\
\end{longtable}
\normalsize

\setlength{\tabcolsep}{4.5pt}
\renewcommand{\arraystretch}{1.2}
\footnotesize
\begin{longtable}{@{}lrrrrrrrrrrrrrrr@{}}
\caption{U.S. -- Posterior summary for scalar parameters (DMP-AP and DMP-LC), by sex and in-sample start year (1979 vs 1989).}
\label{tab:us_posteriors_scalars_all}\\
\toprule
Year & \multicolumn{2}{l}{Parameter} & \multicolumn{5}{c}{Males} & \multicolumn{7}{c}{Females} \\
& & Mean & SD & 5\% & 50\% & 95\% & $n_{\mathrm{eff}}$ & $\widehat{R}$ & Mean & SD & 5\% & 50\% & 95\% & $n_{\mathrm{eff}}$ & $\widehat{R}$ \\
\midrule
\endfirsthead

\toprule
Year & \multicolumn{2}{l}{Parameter} & \multicolumn{5}{c}{Males} & \multicolumn{7}{c}{Females} \\
& & Mean & SD & 5\% & 50\% & 95\% & $n_{\mathrm{eff}}$ & $\widehat{R}$ & Mean & SD & 5\% & 50\% & 95\% & $n_{\mathrm{eff}}$ & $\widehat{R}$ \\
\midrule
\endhead

\midrule
\multicolumn{15}{r}{\small continued on next page}\\
\endfoot

\bottomrule
\endlastfoot

\multicolumn{14}{l}{\hspace{-.08in} \textbf{\underline{DMP-AP}}}\\
1979 & $\nu_0$           & -4.65  & 0.00  & -4.65 & -4.65 & -4.65 & 2674 & 1.00 & -5.18  & 0.00  & -5.18 & -5.18 & -5.18 & 2990 & 1.00\\
& $\sigma_{\delta}$  & 0.70   & 0.12  & 0.54  & 0.68  & 0.91  & 323  & 1.00 & 0.63   & 0.11  & 0.48  & 0.62  & 0.84  & 195  & 1.02 \\
& $\sigma_{\pi}$     & 0.02   & 0.00  & 0.01  & 0.02  & 0.02  & 210  & 1.00 & 0.02   & 0.00  & 0.02  & 0.02  & 0.03  & 264  & 1.00 \\
& $\sigma_{\zeta}$       & 0.34   & 0.03  & 0.30  & 0.34  & 0.39  & 269  & 1.03 & 0.32   & 0.03  & 0.28  & 0.32  & 0.37  & 225  & 1.00 \\
& $\sigma_{\lambda}$       & 0.03   & 0.00  & 0.02  & 0.03  & 0.04  & 249  & 1.00 & 0.01   & 0.00  & 0.01  & 0.01  & 0.02  & 96   & 1.04 \\
& $\phi$             & 483.17 & 12.47 & 462.16& 483.03& 503.70& 217  & 1.00 & 591.35 & 16.80 & 562.09& 591.38& 618.97& 242  & 1.01 \\
\addlinespace

\multicolumn{14}{l}{\hspace{-.08in} \textbf{\underline{DMP-LC}}}\\
1979 & $\nu_0$           & -4.66  & 0.00 & -4.66 & -4.66 & -4.66 & 3172 & 1.00 & -5.18  & 0.00  & -5.18 & -5.18 & -5.18 & 2328 & 1.00 \\
& $\sigma_{\kappa}$  & 0.37   & 0.05 & 0.30  & 0.37  & 0.46  & 273  & 1.01 & 0.31   & 0.04  & 0.25  & 0.31  & 0.39  & 340  & 1.00 \\
& $\sigma_{\zeta}$       & 0.33   & 0.02 & 0.29  & 0.33  & 0.37  & 186  & 1.00 & 0.32   & 0.02  & 0.28  & 0.31  & 0.36  & 290  & 1.02 \\
& $\phi$             & 229.22 & 5.85 & 219.36& 229.40& 238.45& 298  & 1.01 & 416.47 & 10.76 & 398.33& 416.54& 434.21& 275  & 1.01 \\
\addlinespace

\multicolumn{14}{l}{\hspace{-.08in} \textbf{\underline{DMP-AP}}}\\
1989 & $\nu_0$           & -4.72   & 0.00  & -4.72  & -4.72  & -4.72  & 2549 & 1.00 & -5.22  & 0.00  & -5.22 & -5.22 & -5.22 & 2984 & 1.00 \\
& $\sigma_{\delta}$  & 0.71    & 0.12  & 0.54   & 0.70   & 0.94   & 387  & 1.01 & 0.64   & 0.11  & 0.49  & 0.62  & 0.85  & 258  & 1.03 \\
& $\sigma_{\pi}$     & 0.02    & 0.00  & 0.01   & 0.02   & 0.03   & 425  & 1.00 & 0.02   & 0.00  & 0.02  & 0.02  & 0.03  & 378  & 1.01 \\
& $\sigma_{\zeta}$       & 0.35    & 0.03  & 0.31   & 0.35   & 0.40   & 254  & 1.00 & 0.32   & 0.02  & 0.29  & 0.32  & 0.36  & 346  & 1.02 \\
& $\sigma_{\lambda}$       & 0.03    & 0.00  & 0.02   & 0.03   & 0.03   & 286  & 1.02 & 0.02   & 0.00  & 0.01  & 0.02  & 0.02  & 231  & 1.01 \\
& $\phi$             & 615.65  & 21.02 & 579.38 & 615.89 & 649.42 & 510  & 1.00 & 928.29 & 31.49 & 876.75& 928.39& 980.45& 426  & 1.01 \\
\addlinespace

\multicolumn{14}{l}{\hspace{-.08in} \textbf{\underline{DMP-LC}}}\\
1989 & $\nu_0$          & -4.72   & 0.00  & -4.72  & -4.72  & -4.72  & 2842 & 1.00 & -5.23  & 0.00  & -5.23 & -5.23 & -5.22 & 3424 & 1.00 \\
& $\sigma_{\kappa}$ & 0.44    & 0.08  & 0.34   & 0.43   & 0.59   & 491  & 1.01 & 0.31   & 0.05  & 0.24  & 0.30  & 0.40  & 232  & 1.02 \\
& $\sigma_{\zeta}$      & 0.35    & 0.03  & 0.31   & 0.35   & 0.39   & 467  & 1.00 & 0.32   & 0.02  & 0.29  & 0.32  & 0.36  & 291  & 1.00 \\
& $\phi$            & 545.39  & 18.20 & 516.78 & 545.21 & 574.64 & 472  & 1.00 & 805.97 & 28.19 & 759.64& 805.95& 851.91& 273  & 1.00 \\
\addlinespace
\end{longtable}
\normalsize

\end{document}